\documentclass[twocolumn,pre, superscriptaddress, showpacs]{revtex4}
\usepackage{hyperref,xcolor}
\usepackage{amsmath,amssymb}
\usepackage{graphicx}

\newcommand{\imag}{\ensuremath{\mathrm{i}}}
\renewcommand\Im{\ensuremath{\mathrm{Im}\,}}
\renewcommand\Re{\ensuremath{\mathrm{Re}\,}}
\newcommand\Tr{\ensuremath{\mathrm{Tr}\,}}
\newcommand\Dfrtl[1]{\ensuremath{\,\mathrm{d}#1\,}}
\renewcommand\epsilon{\varepsilon}
\newcommand\sign{\ensuremath{\mathrm{sgn}}}
\newcommand\euler[1]{\ensuremath{\mathrm{e}^{#1}}}

\begin{document}
\title{Continuous-Time Quantum Monte Carlo and Maximum Entropy Approach to an 
Imaginary-Time Formulation of Strongly Correlated Steady-State Transport}
\date{\today}
\author{Andreas \surname{Dirks}}
\affiliation{Department of Physics, University of G\"ottingen, D-37077
G\"ottingen, Germany}
\author{Philipp \surname{Werner}}
\affiliation{Institut f\"ur theoretische Physik, ETH Zurich, CH-8093
Z\"urich, Switzerland}
\author{Mark \surname{Jarrell}}
\affiliation{Department of Physics and Astronomy, Louisiana State University,
Baton Rouge, LA 70803, USA}
\author{Thomas \surname{Pruschke}}
\affiliation{Department of Physics, University of G\"ottingen, D-37077
G\"ottingen, Germany}

\begin{abstract} 
Recently Han and Heary proposed an approach to steady-state quantum transport
through mesoscopic structures,  which maps the non-equilibrium problem onto a 
family of auxiliary quantum impurity systems subject to imaginary 
voltages. We employ continuous-time quantum Monte-Carlo solvers to calculate 
accurate imaginary time data for the auxiliary models. The spectral
function is obtained from a maximum entropy analytical continuation 
in both Matsubara frequency and complexified voltage. To enable the analytical continuation 
we construct a kernel which is compatible with the analytical 
structure of the theory. While it remains a formidable task to extract reliable 
spectral functions from this unbiased procedure,
particularly for large voltages, our results indicate that the method in principle yields results in agreement
with those obtained by other methods.  
\end{abstract}


\maketitle
\section{Introduction}
The calculation of steady-state transport properties of open quantum systems 
such as quantum dots is a challenging and unsolved problem. Perturbative 
methods 
\cite{hershfield91prl, hershfield91prb, fujiiueda03}
may be used to study the weak correlation regime, but they fail to 
provide a reliable description of the competition between Kondo- and 
Coulomb-blockade physics in strongly interacting dots \cite{Werner10}. To avoid these
limitations of conventional perturbation theory, various non-perturbative numerical 
approaches have been developed. Time-dependent density-matrix 
renormalization group (tDMRG) calculations \cite{Schmitteckert04, heidrichmeisner09}
 and real-time Monte Carlo (RT-MC) 
approaches \cite{Muehlbacher08, Weiss08,Werner09, Shiro09} try to compute the relaxation into the interacting steady state
after some  switching  of parameters, such as voltage bias or interaction.
While the short-time transients can be very accurately captured with these 
methods \cite{Schmidt08}, the approach to the steady-state may occur on rather long, in the worst
case exponentially large times scales. Due to finite-size effects in the
tDMRG and an exponentially growing sign problem with increasing time in RT-MC, the access
to long times is severely limited in both approaches. Furthermore,
the tDMRG is performed for a finite, closed system; whether a 
relaxation to a reasonable approximation of the interacting steady-state is
guaranteed for some intermediate time scale much smaller than
Poincar\'e's recurrence time is not obvious.
This latter problems may be avoided by numerical renormalization group (NRG) 
\cite{Anders08}
and 
functional renormalization group (fRG) calculations \cite{Rosch2005,Jakobs07, Gezzi07,Pruschke09nato, Schoeller09,Pruschke09},
 which attempt a direct 
description of the non-equilibrium steady state. However, the former 
introduces an artificial discretization and truncation of the spectrum of the
Hamiltonian, which can lead to artifacts in the time evolution.
The fRG, on the other hand, is again
perturbative in nature, and experience up to now shows that it works best
in the extreme non-equilibrium limit \cite{Schoeller09}. 

None of the methods developed so far is able to provide a
complete and reliable description of the model in all parameter 
regimes. More importantly, the most interesting regime, where all relevant energy scales --
voltage, temperature, magnetic field etc.\ -- 
are of the same order as the relevant low-energy scale of the model, is usually
the one which is not accessible. Therefore, the development of new or improved simulation
approaches is a worthwhile and important task. 

Recently, a new and rather unconventional approach to calculate the 
steady-state transport through interacting quantum dots or similar structures
was proposed by Han and Heary \cite{Han07}. Their formalism, which is based on
Hershfield's density operator \cite{Hershfield93}, maps the non-equilibrium
steady-state of the interacting model onto an infinite set of auxiliary 
equilibrium systems, each characterized by some complex voltage. 
The appealing feature of this approach is that powerful methods exist for the 
numerical solution of equilibrium models. The complexification of the voltage 
bias, however, introduces a formidable new problem in the form of an analytical
continuation in the voltage on top of the already challenging analytical 
continuation from Matsubara frequencies to real frequencies. In Ref.~\cite{Han07} 
this double analytical continuation was performed using a phenomenological
formula based on general structures of the self-energy found in second order
perturbation theory.  

The purpose of this study is to explore to what extent an unbiased numerical 
implementation of the method by Han and Heary is feasible. We will address two 
issues: (i) the use of recently developed, accurate continuous-time 
quantum Monte-Carlo
(CT-QMC) algorithms to simulate quantum impurity models as solvers for the effective
equilibrium impurity problems with complex voltage bias; and (ii) the analytical
continuation of Matsubara frequency data via some Maximum Entropy method. In 
particular, we will compare the performance of the weak-coupling \cite{Rubtsov05} and 
hybridization expansion \cite{Werner06} algorithms and
propose a kernel for the Maximum Entropy (ME) procedure which is compatible with
the analytical properties of the Green function. 

The paper is organized as follows. Section~\ref{Han} describes the imaginary-time 
approach to steady state transport by Han and Heary. A brief introduction to the
CT-QMC for equilibrium problems and their suitability for models with complex voltage 
bias follows in section~\ref{QMC}. Section~\ref{MaxEnt} is devoted to the issue
of analytical continuation in the voltage and frequency domain and presents some
results for equilibrium and non-equilibrium situations. We will finish the paper
with a conclusion and outlook in section~\ref{Conclusion}.

\section{Imaginary-Time Formulation of Steady-State Transport}
\label{Han}

We briefly review the imaginary-time formulation of steady-state 
transport through an interacting quantum dot proposed by Han and Heary \cite{Han07}, which is
based on the work of Hershfield \cite{Hershfield93}. 

\subsection{Physical Model}
We consider a spin-degenerate, single-level 
quantum dot attached to two non-interacting fermionic leads. This system can be
described by the Single-Impurity Anderson Model with Hamiltonian ($e=\hbar=1$)
\begin{eqnarray}
H &=& H_0 +H_\text{int} ,\phantom{\Big)} 
\label{eq:PhysicalHamiltonian}
\\
H_0 &=& \sum_{\alpha k\sigma} \epsilon_{\alpha k\sigma} c^\dagger_{\alpha k\sigma}
 c^{\phantom{\dagger}}_{\alpha k\sigma} + \sum_\sigma V_G d_\sigma^\dagger d_\sigma^{\phantom{\dagger}}\nonumber\\
&&  + \sum_{\alpha k\sigma} 
\left(V_{\alpha k\sigma}c^\dagger_{\alpha k\sigma}d_\sigma^{\phantom{\dagger}} +
V_{\alpha k\sigma}^*d^\dagger_\sigma c_{\alpha k\sigma}^{\phantom{\dagger}} \right),\\
H_\text{int} &= & U \left(n_\uparrow - \frac{1}{2}\right) \left(n_\downarrow - \frac{1}{2}\right),
\end{eqnarray}
where $\alpha=-1$ and $\alpha=+1$ label the left and right reservoirs,
respectively. The index $k$ denotes the wave-vector of the
lead states and $\sigma$ the spin quantum number. A gate voltage $V_G$ may be applied
to shift the dot energy level position relative to the particle-hole symmetric
configuration $V_G=0$. 

To keep things simple, we assume a $k$-independent hybridization  $V_{\alpha k\sigma}= V/\sqrt{2}$
and consider the wide-band limit for the dispersion of the leads.
We then end up with a bare level broadening
$\Gamma=\Gamma_L+\Gamma_R$, $\Gamma_\alpha = \pi |V|^2 N_F/2$, where $N_F$ denotes 
the density of states of the leads at the Fermi energy.

In the case of non-equilibrium steady-state transport, the leads are supposed 
to be unaffected by the current flowing through the dot and characterized by 
free Fermion correlators  
\begin{equation}
\langle c^{\dagger}_{\alpha p \sigma}c^{\phantom{\dagger}}_{\beta p'\sigma'}\rangle=\delta_{\alpha,\beta}\delta_{p,p'}\delta_{\sigma,\sigma^{'}}f_{\beta_\alpha}(\epsilon^\alpha_{p,\sigma}-\mu_\alpha),
\end{equation}
with $f_\beta(x)=(e^{\beta x}+1)^{-1}$ the Fermi distribution function for inverse temperature $\beta$ 
and $\mu_\alpha$ the value of the chemical potential for lead $\alpha$.
We restrict ourselves to the case where the inverse temperatures of the left and right lead are
the same, $\beta_L = \beta_R = \beta$, and symmetrically applied voltage bias, $\mu_L=-\mu_R$.
 The bias voltage is denoted by $\Phi = \mu_L - \mu_R$.

\subsection{The $Y$-Operator}
In Ref.~\onlinecite{Hershfield93}, Hershfield introduced a Hermitian operator $Y$ 
by means of which the non-equilibrium, steady-state expectation value of a 
local observable $A$ may be written as
\begin{equation}
\langle A \rangle
= 
\frac{\Tr \euler{-\beta(H-\Phi Y)} A}{\Tr \euler{-\beta(H-\Phi Y)}}.
\label{eq:expectation_hershfield}
\end{equation}
The above expectation value is of the form $\langle A \rangle = \Tr \rho A/\Tr \rho$, and hence resembles the equilibrium expression. 
Under certain assumptions involving the non-trivial exchange of limiting 
procedures, the operator $Y$ can be expressed as
\begin{equation}
Y = \sum_{\alpha k\sigma}\frac{\alpha}{2}\psi^\dagger_{\alpha k\sigma}\psi^{\phantom{\dagger}}_{\alpha k\sigma},
\end{equation}
where the scattering states $\psi_{\alpha k \sigma}$ are related to the bare
conduction states $c_{\alpha k \sigma}$ by the second-quantized 
Lippmann-Schwinger equation \cite {hanprb07}
\begin{equation}
\psi^\dagger_{\alpha k \sigma} =  c^\dagger_{\alpha k\sigma} + \frac{1}
{\epsilon_{\alpha k\sigma} - \mathcal{L} + \mathrm{i}\eta} \mathcal{L}_V
c^\dagger_{\alpha k\sigma}.
\label{LSeq}
\end{equation}
The Liouvillians are defined as $\mathcal{L} = [H,\cdot]$ and
$\mathcal{L}_V=[H_V,\cdot]$,
with $H_V =
\sum_{\alpha k\sigma} 
(V_{\alpha k\sigma}c^\dagger_{\alpha k\sigma}d_\sigma^{\phantom{\dagger}} +
\text{h.c.})$
the hybridization part of the Hamiltonian. 
The ``$\cdot$" denotes the operators after $\mathcal L$, and the 
fraction in Eq.~(\ref{LSeq}) denotes the corresponding geometric series in $\mathcal{L}$, 
i.e.~a series of iterated commutators with $H$.\par
For $U\neq 0$ it is impossible to calculate an explicit expression 
for the $Y$-operator. 
More importantly, although $H-\Phi Y$ looks like an effective Hamiltonian for 
the system, it cannot be used to define a consistent description of 
imaginary-time and real-time dynamics. The real-time dynamics is always controlled by $H$
alone, but $H$ and $H-\Phi Y$ will in general have a different spectrum. Therefore, 
the analytically continued imaginary-time dynamics does not reproduce the
real-time dynamics.

\subsection{Imaginary Voltages}
Since $H-\Phi Y$ does not yield the correct real-time dynamics, Han and Heary \cite{Han07}
introduce an additional trick. Starting with a fully established 
non-interacting steady-state ensemble at time $t=0$, the fully interacting 
steady state is formally reached 
by propagating the system to 
$t=+\infty$. In a path integral representation the expectation value for an observable $A$  becomes 
\begin{equation}\label{eq:true_t_evolution}
\langle A \rangle 
= 
\left\langle 
\int\mathcal{D}[\psi^\dagger, \psi^{\phantom{\dagger}}]
A(\{\psi_{\alpha k \sigma}^\dagger(0), \psi_{\alpha k \sigma}^{\phantom{\dagger}}(0)\})
\euler{\imag\int_0^\infty L(t)\Dfrtl t}
\right\rangle_0.
\end{equation}
Here, the average $\langle\cdot\rangle_0$ is performed using
Eq.~\eqref{eq:expectation_hershfield} with $H\to H_0$ and $Y\to Y_0$,
where $Y_0$ can be explicitly constructed using non-interacting scattering
states.
It was argued in Ref.~\onlinecite{Han07} that the time evolution via $H$
maps the non-interacting scattering states to the interacting ones and
the Lagrangian for the real-time evolution reads 
\begin{equation}
L(t) = \sum_{\alpha k \sigma} \psi^\dagger_{\alpha k \sigma}(t)(\imag \partial_t -
\epsilon_{\alpha k \sigma})\psi_{\alpha k\sigma}(t).
\end{equation}
Aiming at a description which yields $\euler{\imag H (t'-t)}$ as real-time evolution operator
for $t\to t'$
and $\euler{-(\tau'-\tau)(H-\Phi Y)}$ as imaginary-time evolution operator
for $-\imag \tau \to -\imag \tau'$, the Lagrangian is
reexpressed with respect to the spectrum of $H-\Phi Y$, 
$\tilde\epsilon_{\alpha k \sigma} = \epsilon_{\alpha k \sigma} - \alpha \Phi
/2$. Statistical expectation values take a form analogous to equilibrium
expectation values, with a uniform Fermi level $\tilde\epsilon_{\alpha k\sigma} = 0$.
Due to the discrepancy between $H$ and $H-\Phi Y$, the real-time Lagrangian transforms to 
$L(t) = \sum_{\alpha k \sigma} \psi_{\alpha k \sigma}^\dagger(t)
(\imag \partial_t - \tilde\epsilon_{\alpha k \sigma}-\alpha \Phi/2)
\psi_{\alpha k \sigma}^{\phantom{\dagger}}(t)$, so the effective Fermi levels of left and right
leads have different time evolution rates. These rates can be factored out as
time-dependent phase factors of the Grassmann fields by 
introducing new field variables 
$\tilde\psi_{\alpha k \sigma} (t) = \euler{\imag\alpha \Phi t /2} \psi_{\alpha k
\sigma} (t)$. The extra time evolution rate is
generated by $\imag\partial_t$ acting on the phase factor, and thus
$L(t) = \sum_{\alpha k \sigma} \tilde\psi_{\alpha k \sigma}^\dagger(t)
(\imag \partial_t - \tilde\epsilon_{\alpha k \sigma})
\tilde\psi_{\alpha k \sigma}^{\phantom{\dagger}}(t)$ describes the correct time evolution.

To obtain a Matsubara-like theory, the fields $\tilde \psi$ are now
Wick rotated, $\tilde \psi(t) \to \tilde \psi (-\imag\tau)$. However,
under the replacement $t\to -\imag \tau$ the exponential factor becomes 
$\euler{\alpha \Phi\tau /2}$, which means that it diverges
as $\tau\to\infty$ and decays as $\tau\to-\infty$.
To circumvent this problem, Han and Heary introduce a second analytic 
continuation to ensure Matsubara's periodic boundary conditions and thereby 
obtain a well-defined effective equilibrium system. This is achieved by
complexifying the voltage occurring in the
extra time evolution rate according to
$\Phi \to\imag \varphi_m$, $m\in \mathbb{Z}$. For the
particular choice $\varphi_m=4\pi m / \beta$
the Matsubara boundary conditions are conserved \cite{Han07}.

\subsection{Effective Action}
The final result of these manipulations is that both the Lagrangian
and the fields now have their time evolution with respect to the
effective
equilibrium Hamiltonian $K=H - (\Phi -\imag \varphi_m) Y$.
In a perturbative expansion around the non-interacting limit, one may then 
switch to the interaction picture with respect to the non-interacting effective
Hamiltonian $K_0=H_0 - (\Phi -\imag \varphi_m) Y_0$. As before, $Y_0$ is Hershfield's
boundary condition operator for the corresponding fully established
non-interacting steady state, for which an explicit expression can be given.

We may now proceed along the usual lines and integrate
out the conduction electron degrees of freedom to obtain an
effective action 
\begin{widetext}
\begin{equation}
S_\text{eff} = \sum_\sigma\iint_0^\beta \Dfrtl \tau \Dfrtl{\tau'}
d_{\sigma}^\dagger (\tau') G_{0\sigma}^{-1}(\tau',\tau)
d_{\sigma}^{\phantom{\dagger}}(\tau) 
+
U \int_0^\beta \Dfrtl\tau\left(
                     d_{\downarrow}^\dagger(\tau)
                     d_{\downarrow}^{\phantom{\dagger}}(\tau)-\frac{1}{2}\right)\left(
                     d_{\uparrow}^\dagger(\tau)
                     d_{\uparrow}^{\phantom{\dagger}}(\tau)-\frac{1}{2}\right)\;\;
\label{action}
\end{equation}
\end{widetext}
for the electrons on the dot. As we are by construction in the stationary state,
the bare dot Green's function $G_{0\sigma}(\tau',\tau)$ appearing in the
quadratic term in the action (\ref{action}) depends on the time difference only.
We therefore may perform a Fourier transform to fermionic Matsubara frequencies and
find the form \cite{Han07}
\begin{equation}
G_{0,mn}
= \sum_{\alpha = \pm 1}
\frac{1/2}{\imag\omega_n - \frac{\alpha}{2}(\imag\varphi_m-\Phi) - \epsilon_d + \imag
\Gamma^{(\alpha)}_{mn}
}\;\;,
\label{eq:bare_gf}
\end{equation}
with $G_{0,mn} := G_0 (\imag \varphi_m, \imag\omega_n)$,
$\Gamma^{(\alpha)}_{mn} := \Gamma\sign(\omega_n - \alpha\varphi_m/2)$,
and $\epsilon_d = V_G$. 

The desired Green's function for the stationary state of the interacting 
system is finally obtained by solving the quantum impurity problem for each 
$\imag\varphi_m$, $m\in\mathbb{Z}$, performing the analytical continuation
$\imag\varphi_m\to z_\varphi$ and evaluating the resulting expression at the 
physical voltage $z_\varphi = \Phi$. 

Although the preceding discussion seems to be based
on simple manipulations of the functional integral, one has to show formally
the equivalence of the complexified auxiliary equilibrium time-evolution based
on the action (\ref{action}) and the actual physical time evolution with 
respect to $H$ as given by (\ref{eq:true_t_evolution}) after the analytical
continuation $\imag\varphi_m\to \Phi$ in the former. Up to now such a formal 
proof is still lacking, only an argument based on the inspection of the
contributions to perturbation expansion has been put forward \cite{Han07}.
It is therefore interesting to see if an unbiased numerical implementation of this formalism is possible and produces physically meaningful results.

\section{Continuous-Time Quantum Monte Carlo}
\label{QMC}

In order to compute the self-energy from action (\ref{action}) as a function of
Matsubara frequency we employ continuous-time Monte Carlo (CT-QMC) solvers. The
continuous-time Monte Carlo technique in the weak-coupling \cite{Rubtsov05} and
hybridization expansion \cite{Werner06} formulation has been discussed in considerable 
detail in the literature and we will present here merely a short summary of the 
formalism. The idea is to expand the partition function $Z=\Tr[e^{-\beta H}]$ into 
a series of diagrams, and to sample (collections of) these diagrams by a Monte 
Carlo procedure. We split the Hamiltonian $H$ of the impurity model into two 
parts, $H_1$ and $H_2=H-H_1$, and employ an interaction representation in which 
the time evolution of operators is given by $H_1$: $O(\tau)=e^{\tau H_1}O e^{-\tau H_1}$. 
In this interaction representation, the partition function can be expressed as 
a time ordered exponential, which is then expanded into powers of $H_2$,
\begin{widetext}
\begin{equation}
Z=\Tr \Big[e^{-\beta H_1} Te^{-\int_0^\beta d\tau H_2(\tau)} \Big]
=
\sum_{n=0}^\infty \int_0^\beta d\tau_1\cdots \int_{\tau_{n-1}}^\beta d\tau_n \Tr\Big[ e^{-(\beta-\tau_n)H_1}(-H_2) \cdots e^{-(\tau_2-\tau_1)H_1}(-H_2)e^{-\tau_1H_1}\Big].
\label{Z_interaction_picture}
\end{equation}
Equation~(\ref{Z_interaction_picture}) represents the partition function as a sum over Monte Carlo configurations $c=\{\tau_1<\ldots<\tau_n\}$; $n=0$, $1$, $\ldots$, $\tau_i\in[0,\beta)$ with weight
\begin{equation}
w_c=\Tr\Big[ e^{-(\beta-\tau_n)H_1}(-H_2)\cdots e^{-(\tau_2-\tau_1)H_1}(-H_2)e^{-\tau_1H_1}\Big]d\tau^n.
\label{weight0}
\end{equation}
\end{widetext}
Two types of expansions have been considered. In the weak-coupling approach
\cite{Rubtsov05} the partition function is expanded into powers of the
interaction, $H_2=H_\text{int}$, while the time evolution between operators
is given by the quadratic part of the Hamiltonian, $H_1=H_0$. The Monte Carlo
configuration becomes a collection of interaction vertices on the imaginary
time interval and the weight (\ref{weight0}) evaluates to
\begin{equation}
w_c^\text{weak}=(-U)^n\det \Big[G_0-\frac{1}{2}I\Big]d\tau^n.
\label{weight}
\end{equation}
Here $(G_0)_{ij}=G_0(\tau_i-\tau_j)$ is an $n\times n$ matrix whose elements are 
noninteracting Green functions evaluated at all time intervals defined by the vertex 
positions. Note that in the case of half filling of interest here, only even perturbation 
orders appear in the expansion. Away from half-filling, odd perturbation orders become 
relevant and Ising-type auxiliary fields must be introduced to avoid or reduce the 
sign problem. We will in this paper employ the continuous-time auxiliary field algorithm 
described in Ref.~\cite{Gull08_ctaux}, which for models with density-density interactions 
and an appropriate choice of parameters 
is equivalent to the weak-coupling algorithm \cite{Mikelsons09}.

The alternative approach is the hybridization expansion \cite{Werner06} where
the partition function is expanded in powers of the hybridization term,
$$
H_2=\sum\limits_{\alpha k \sigma} (V_{\alpha k \sigma}c^\dagger_{\alpha k
\sigma}d_\sigma^{\phantom{\dagger}}+ \text{h.~c.}),
$$
while the time evolution between operators is given by the impurity plus bath part 
of the Hamiltonian. This time evolution no longer couples the impurity and the 
bath. It therefore becomes possible to integrate out the bath degrees of freedom
analytically to obtain
\begin{widetext}
\begin{eqnarray}
w_{\tilde c}&=&Z_\text{bath}
\Tr_\text{loc} \Big[e^{-\beta H_\text{loc}} T 
\psi_{\alpha_{n}}(\tau_n)\psi^\dagger_{\alpha_n'}(\tau_{n}')
\cdots
\psi_{\alpha_1}(\tau_1)\psi^\dagger_{\alpha_1'}(\tau_1') 
\Big] \nonumber\\
&& 
\times \det M^{-1}(\{\tau_1, \alpha_1\},\ldots,\{\tau_{n},\alpha_n\}; \{\tau_1',\alpha_1'\},\ldots,\{\tau_{n}',\alpha_n'\}) (d\tau)^{2n}.
\label{weight_strong}
\end{eqnarray}
\end{widetext}
The configurations $\tilde c$ are now collections of $n$ time arguments $\tau_1<\ldots<\tau_n$ 
corresponding to annihilation operators with flavor indices $\alpha_1, \ldots,\alpha_n$ 
and $n$ time arguments $\tau_1'<\ldots<\tau_n'$ corresponding to creation operators 
with flavor indices $\alpha_1', \ldots, \alpha_n'$. The element $i,j$ of the matrix 
$M^{-1}$ is given by the hybridization function $F_{\alpha_i',\alpha_j}(\tau_i'-\tau_j)$, 
which is defined in terms of the hybridization parameters $V^{\alpha, \alpha'}_p$ 
and the bath energy levels $\epsilon^\alpha_p$ \cite{Werner06Kondo}. 
In a model with density-density interactions only, one can separate the operators
 according to flavors, which leads to the so-called segment representation \cite{Werner06}. 
This segment representation allows a simple and efficient evaluation of the trace 
over the impurity states in Eq.~(\ref{weight_strong}). 
\subsection{Implementation}
The implementation of the weak-coupling CT-QMC for the action 
\eqref{action} is straightforward. The non-interacting Green's function 
\eqref{eq:bare_gf} is being Fourier-transformed and the resulting $G_{0,mn}(\tau)$ inserted into Eq.~\eqref{weight}. 

The implementation of the hybridization approach is more subtle, 
as -- except in the equilibrium limit $\Phi=0$, $\imag\varphi_m=0$ -- 
the hybridization function $F_{\alpha_i',\alpha_j}(\tau_i'-\tau_j)$
which appears in the action 
\eqref{weight_strong} 
lacks a physical meaning, because it is not directly related to the hopping
amplitudes $V$ in the physical Hamiltonian \eqref{eq:PhysicalHamiltonian}. 
However, the hybridization function is implicitly defined by rewriting the
effective action \eqref{action} as \cite{Werner06} 
$S_\text{eff} = S_{F} + S_{\text{loc}}$, with
$
S_F = -\sum_\sigma\iint_0^\beta \Dfrtl{\tau}\Dfrtl{\tau'}
d_\sigma(\tau) F(\tau-\tau') d_\sigma^\dagger(\tau')$
and
$S_\text{loc} = 
-\int_0^\beta \Dfrtl{\tau} (\sum_\sigma\epsilon_d d^\dagger_\sigma d_\sigma
- U d_\uparrow^\dagger d_\downarrow^\dagger d_\uparrow d_\downarrow )
$. Consequently, the hybridization function
can be constructed from \eqref{eq:bare_gf} as
\begin{eqnarray}
F(-\imag\omega_n) &=&
\imag \omega_n - \epsilon_d - G_0(\imag\varphi_m, \imag
\omega_n)^{-1}\;\;,\\
\label{eq:hybfunction_freqspace}
F(\tau) &=& \frac{1}{\beta}\sum_{n=-\infty}^\infty  \euler {-\imag\omega_n \tau}F(\imag\omega_n).
\label{eq:hybfunction_timespace}
\end{eqnarray}
After straightforward algebraic manipulation, we obtain
\begin{widetext}
\begin{equation}
\label{eq:F_explicit}
F(\imag\omega_n) =
\frac{
\prod\limits_{\alpha=\pm 1} [\imag\Gamma\sign(\omega_n - \frac{\alpha}{2}\varphi_m)]
+
\sum\limits_{\alpha=\pm 1} [\imag\omega_n - \epsilon_d + \alpha
(\imag\varphi_m-\Phi)] \frac{\imag\Gamma}{2} \sign(\omega_n -\frac{\alpha}{2}\varphi_m)
- \left(\frac{\imag \varphi_m - \Phi}{2}\right)^2
}
{
\imag\omega_n - \epsilon_d + \frac{\imag \Gamma}{2} \sum\limits_{\alpha = \pm 1}
\sign(\omega_n - \frac{\alpha}{2} \varphi_m)
}\;\;.
\end{equation}
\end{widetext}
Note that the expression $\imag\Gamma\sign(\omega_n-\alpha
\varphi_m/2)$ emerges from imposing the wide-band limit for the leads. 
The hybridization approach is only able to cope with finite bands, because in  
the limit of infinitely wide bands of constant DOS, the expansion order diverges.
The $\sign$-function must therefore be replaced by a sufficiently well-behaved
function corresponding to a finite bandwidth and thus decaying rapidly enough for 
large frequencies $\omega_n$.

The high-frequency behavior of expression \eqref{eq:F_explicit}
is given by
\begin{eqnarray}
F(\imag\omega_n) &\stackrel{|\omega_n|\to\infty}{\to}&
\sum\limits_{\alpha=\pm1}\frac{\imag\Gamma}{2}\sign\left(\omega_n-\frac{\alpha}{2}\varphi_m\right)+\frac{c_1}{\imag\omega_n}\nonumber\\
&=:&\hat{F}(\imag\omega_n)+\frac{c_1}{\imag\omega_n}\nonumber\\
c_1 &=& -\Gamma^2- \left(\frac{\imag \varphi_m - \Phi}{2}\right)^2\;\;,
\label{eq:shifthyb}
\end{eqnarray}
which means that the numerical evaluation of
Eq.~\eqref{eq:hybfunction_timespace} requires some care.
Conventionally, one regularizes the sum by analytically evaluating the dangerous
parts and then numerically calculating the difference between the full 
function and the problematic parts, i.e.\
$$
\Delta F(\tau):=
\frac{1}{\beta}\sum\limits_{n=-\infty}^\infty\left[F(\imag\omega_n)
-\left(\hat{F}(\imag\omega_n)+\frac{c_1}{\imag\omega_n}\right)\right]\euler{-\imag\omega_n\tau}\;\;.
$$

The leading order high-frequency tail $c_1/(\imag \omega_n)$ 
results  in a constant shift $-c_1/2$ in $F(\tau)$, $0<\tau<\beta$.
The first term 
$$
\hat{F}(\imag\omega_n):=\sum_{\alpha = \pm 1} \frac{\imag \Gamma}{2}
\sign\left(\omega_n - \alpha \frac{\varphi_m}{2}\right)
$$
in the high-frequency expansion 
yields
\begin{equation}
\hat{F}(\tau)=
\frac{1}{\beta}
\sum_{n=-\infty}^\infty
\hat F(\imag\omega_n)
\euler {-\imag\omega_n\tau}
= 
\frac{\Gamma}{\beta}
\frac{\cos (\varphi_m \tau / 2)}{\sin (\pi \tau /\beta)}
\label{eq:exactfourierwblcontrib}
\end{equation}
and diverges for $\tau\to0$ and $\tau\to\beta$. These divergences are a direct 
consequence of the wide-band limit, i.e.\ we need to regularize them in order to 
be able to use the hybridization expansion algorithm.
This regularization is introduced by cutting the divergences
with a sufficiently large cutoff parameter $F_\text{cut}$,
i.e.\ we use
$$
F(\tau) = \Delta F(\tau)-\frac{c_1}{2}+\min\left(\hat{F}(\tau),F_\text{cut}\right)\;\;.
$$
In practice, the value $F_\text{cut} = 10^{4}$ was used.
The contribution $\Delta F$ is Fourier transformed easily by accumulating the
series numerically. \par
Note that the term $\hat F$ has, besides the additional oscillations from the
cosine modulation in Eq.~\eqref{eq:exactfourierwblcontrib}, the same structure as 
in the plain equilibrium Anderson model, where $F(\tau) =
\frac{\Gamma}{\beta}(\sin(\tau/\beta))^{-1}$.
We will therefore illustrate the properties of the quantity
\begin{equation}
\tilde F(\tau) = F(\tau) - \hat F(\tau)
\label{eq:defFtilde}
\end{equation}
in the following section.
\par
\begin{figure}
\includegraphics[width=0.49\textwidth]{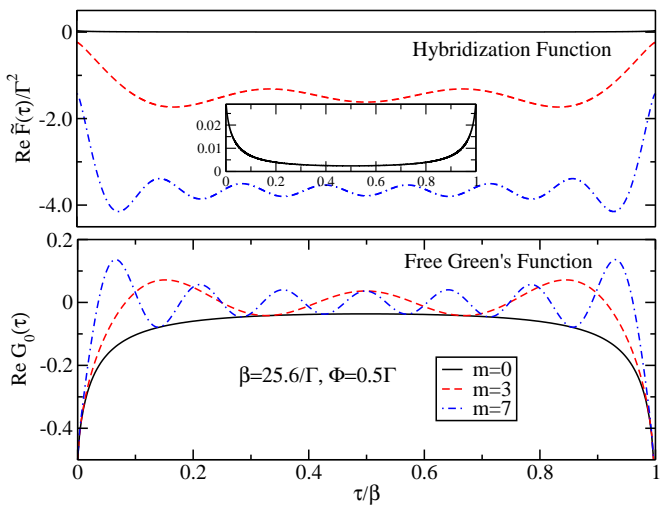}
\caption{(color online) Imaginary-time data used as input for the CT-QMC solvers for different
values of the imaginary voltage $\varphi_m$. 
The upper panel shows the non-trivial contribution $\tilde F(\tau)$, 
Eq.~\eqref{eq:defFtilde} to
the hybridization function $F(\tau)$, the lower panel shows the imaginary-time 
Green's function $G_0$.
Raising $\varphi_m$ leads to increasingly oscillating imaginary-time 
Green's functions and hybridization functions. 
The oscillations need to be resolved well by the QMC solver in order to
guarantee an unbiased solution.
As implied by 
Eq.~\eqref{eq:shifthyb} a strong negative shift $-c_1/2$ occurs
in the hybridization function when sweeping through the region $\varphi_m\gg \Phi$.
The imaginary parts $\Im F(\tau)$ and $\Im G_0(\tau)$ are small and also show
oscillations.
}
\label{fig:imagtimedata}
\end{figure}
\subsection{Imaginary-Time Data}
Typical input data for both, the weak-coupling and the strong-coupling
approach, are shown in Fig.~\ref{fig:imagtimedata}.
With increasing imaginary voltage $\varphi_m$, oscillations with $m$ nodes 
occur in both, the imaginary-time Green's function and the hybridization function.
Moreover, the shift \eqref{eq:shifthyb} grows quadratically, introducing a 
strong shift of the hybridization function towards negative values.

The strongly oscillatory behavior for large $\varphi_m$ makes a correspondingly 
fine resolution of the imaginary-time interval necessary. 
In a standard Hirsch-Fye algorithm \cite{Hirsch86}, the interval $[0,\beta)$ has to be
represented by a comparatively small and fixed number of equidistant mesh
points, i.e.\  these oscillations cannot be adequately resolved.
This limitation does not apply to CT-QMC, and it is hence the method of
choice to access also large $\varphi_m$.

\subsection{Phase Problem}

\begin{figure}
\includegraphics[width=0.49\textwidth]{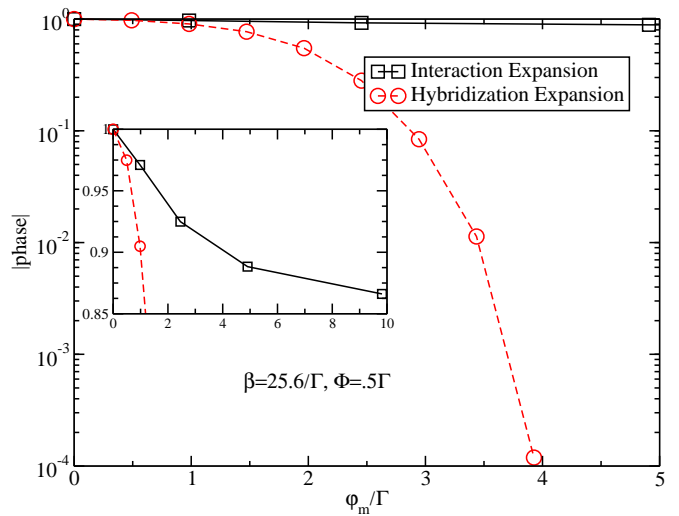}
\caption{(color online) Absolute values of the average sampling weight phases
$|\langle w_c/|w_c| \rangle|$
and $|\langle w_{\tilde c}/|w_{\tilde c}| \rangle |$
(Eqs.~\eqref{weight} and \eqref{weight_strong})
of the weak-coupling (solid lines) and the
strong-coupling (dashed lines) solver, respectively, as a function of the
imaginary
voltage. On a logarithmic scale, the average phase decays faster than linearly
for the strong-coupling approach when $\varphi_m$ is increased.
No strong dependence on $\varphi_m$ is found for the weak-coupling algorithm.
}
\label{fig:phase}
\end{figure}

In contrast to the equilibrium case, complex sampling weights 
$w_c=e^{i\gamma}|w_c|$ are obtained in 
both the weak-coupling and strong coupling formulation. As usual, one uses
the modulus $|w_c|$ of the weight to determine the acceptance probability, 
while the
phase $e^{i\gamma}$ has to be treated as additional observable. Usually,
such an approach leads to a sign problem and severely limits the applicability
of the Monte-Carlo simulations. Therefore, we must anticipate a
generalized sign problem, i.e.\ $\langle e^{i\gamma}\rangle\to0$ exponentially
or worse.
The situation is especially problematic for the hybridization expansion due to 
the additional shift \eqref{eq:shifthyb} towards negative values. Indeed,
as illustrated in Fig.~\ref{fig:phase} the sign problem becomes increasingly
severe with increasing imaginary voltage $\varphi_m$, 
limiting this algorithm to small $\varphi_m$. 
From Fig.\ \ref{fig:phase} it also becomes clear that the sign problem in the
weak-coupling CT-QMC simulations 
is much milder and this approach allows us to simulate impurity models with 
large $\varphi_m$. 

\begin{figure}
\includegraphics[width=0.49\textwidth]{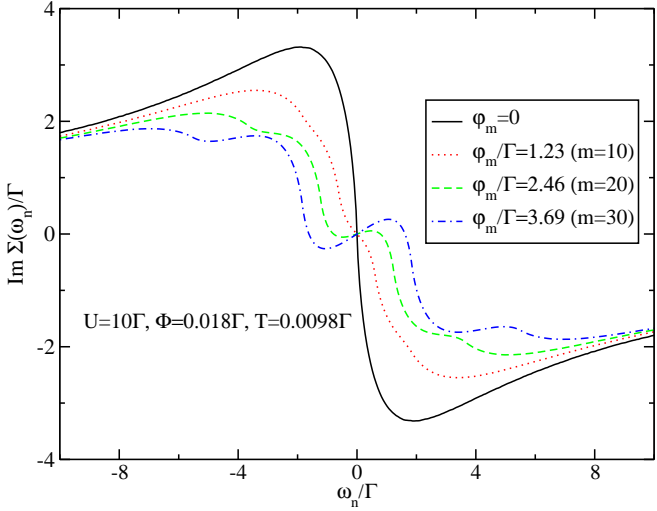}
\caption{(color online) Imaginary part of the impurity self-energy obtained with the
weak-coupling CT-QMC solver for $V_G=0$, $U/\Gamma=10$, $T/\Gamma=0.0098$ and
$\Phi/\Gamma=0.018$ . The equilibrium
Kondo scale here is $T_{\rm K}/\Gamma\approx 0.018$. We easily obtain
high-quality data for all
values $m=10$, $20$ and $30$ of the complexified voltage, even in 
this most challenging parameter regime $T_{\rm K}\ll\Gamma$, $\Phi\approx T_{\rm K}$ and $T\approx T_{\rm K}/2$.
Each $m$-value was run on a single Intel Xeon(R) E5345 CPU for approx.~24 hours, so 
the data were obtained with relatively moderate computational effort.
}
\label{fig:selfenergy}
\end{figure}

To demonstrate the quality of the imaginary-time data which can be obtained
with the weak-coupling CT-QMC method, we show in Fig.~(\ref{fig:selfenergy})
the imaginary part of the Matsubara axis self-energy computed for
$U=10\Gamma$, $\Phi=0.018\Gamma$, $T/\Gamma=0.0098$ 
and $\varphi_m=0$ ($m=0$), $\varphi_m/\Gamma=1.23$ ($m=10$), $\varphi_m/\Gamma=2.46$ ($m=20$), and $\varphi_m/\Gamma=3.69$ ($m=30$). 
The equilibrium Kondo
temperature for this parameter set is $T_{\rm K}/\Gamma \approx 0.018\ll1$,
i.e.\ we are reasonably deep in the Kondo regime of the Anderson model.
Moreover, the values for $\Phi$ and $T$ are such that 
$T\approx T_{\rm K}/2$ and $\Phi\approx T_{\rm K}$, i.e.\ precisely in the parameter
region which is hard or impossible to access for other methods.
Even for large complex
voltage the accuracy of the numerical data is very good
(error bars on the order of the line width) for both
small and large Matsubara frequencies. 
In contrast to the results presented in Ref.\ \onlinecite{Han07},
which are based on discrete-time Hirsch-Fye simulations,
no discontinuities are observed for $\omega_n \approx \pm \varphi_m/2$ in the
CT-QMC data. We note, however, that a recent preprint \cite{hanpreprint} reports a trick by use of which 
this issue could be resolved within the discrete-time formalism.

\section{Analytic Continuation}

\subsection{Analytic Structure}

\begin{figure}
\includegraphics[width=0.49\textwidth]{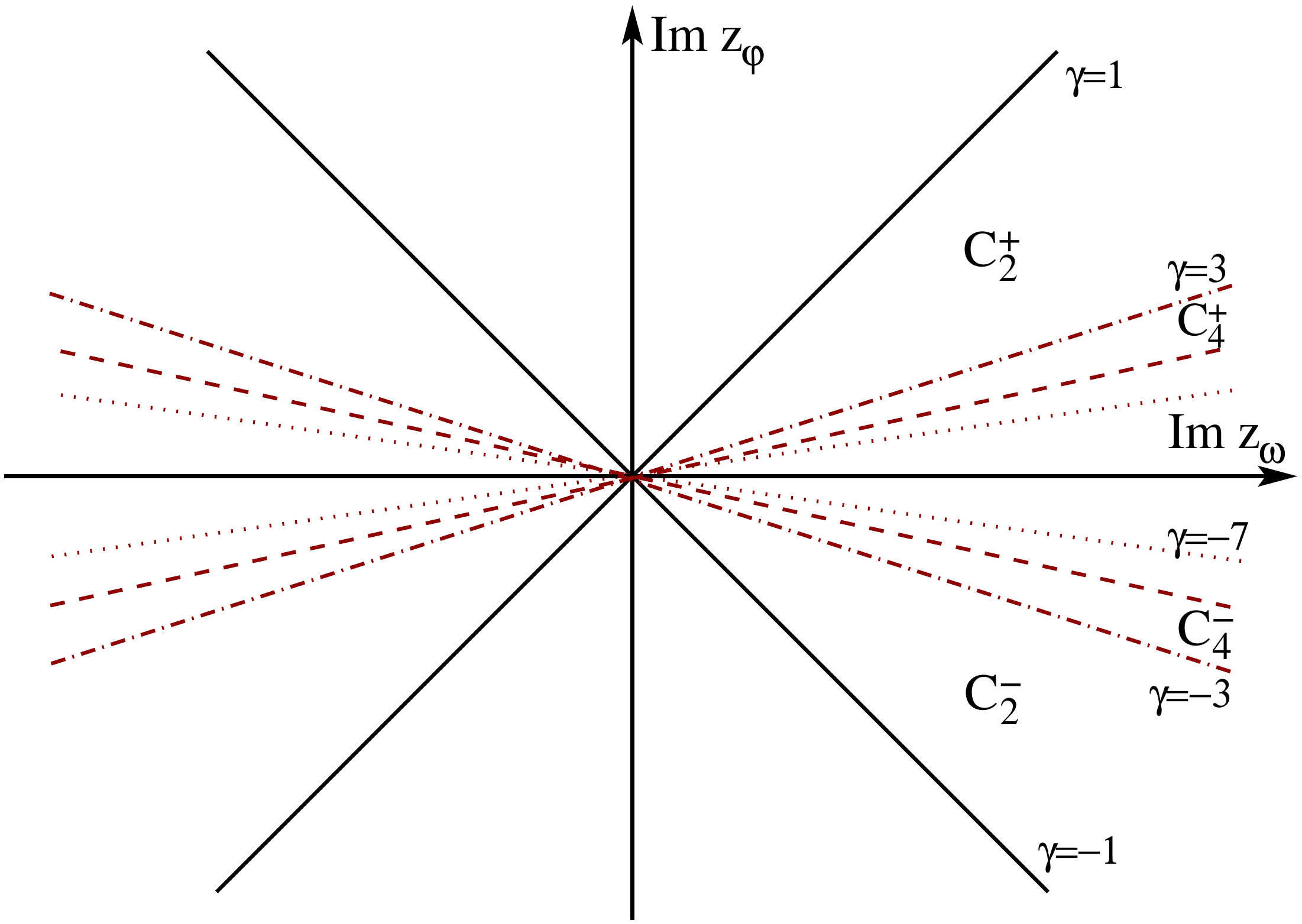}
\caption{(color online) Geometric structure of the complex space carrying the two-variable 
Green's function $G(z_\varphi,z_\omega)$.
Branch cuts occur for $\Im z_\varphi = \frac{2}{\gamma}
\Im z_\omega$, with $\gamma = \pm 1$ (solid lines), for $U=0$,
but also at $\gamma = \pm 3$
(dash-dotted lines), $\gamma = \pm 5$ (dashed lines), $\gamma = \pm 7$ 
(dotted lines), and so on, for $U\ne 0$.
Concentrating on the retarded sector of the Green's function, $\Im z_\omega>0$,
we introduce the cones $C^{\pm}_{\nu}$ bounded by the branch cuts with
imaginary-part ratios $\frac{2}{\nu-1}$ and $\frac{2}{\nu+1}$.
Adding the real subspaces $(\Re z_\varphi, \Re z_\omega)$, the tubular cones
$T^{C^{\pm}_{\nu}} = \mathbb{R}^2 + \imag C^{\pm}_{\nu}$ are obtained as domains of holomorphy.
}
\label{fig:cones}
\end{figure}

As noted in Ref.~\cite{Han07}, at finite interaction, branch cuts occur for
$\Im z_\omega = \frac{\gamma}{2}\Im z_\varphi$ ($\gamma$ odd) in 
the complexified Green's function $G(\imag \varphi_m\to z_\varphi,\imag\omega_n \to z_\omega)$.
Introducing the complex vector variable $\underline z = (z_\varphi, z_\omega)$
we hence assume the Green's function to be holomorphic as a
function of two complex variables 
in domains 
$T^{C^s_\nu} := \mathbb{R}^2 + \imag C^s_\nu$,
where for $\nu\in 2\mathbb{Z}$ 
\begin{equation*}
C^s_\nu := 
\left\{
\begin{pmatrix}
a \\ b
\end{pmatrix}
\in \mathbb{R}^2: s a >
0\,\wedge\,\frac{\nu-1}{2}|a| < b < \frac{\nu+1}{2}|a| \right\}
\end{equation*}
are the cones emerging from the branch cut condition for positive
($s=+1$) or negative ($s=-1$) imaginary voltages (see illustration in Fig.~\ref{fig:cones}). 
Note that domains like $T^{C^s_\nu}$ are well-known objects in the theory of 
functions of 
several complex variables and are called tubular cone domains. 
For a good introduction see, e.~g.,  Ref.~\cite{encyclMathSciences}.

In Ref.~\cite{Han07} this structure is described by the 
Cauchy representation
\begin{equation}
\Sigma(\imag \varphi_m, \imag \omega_n) \approx
\sum_{\gamma \in 2 \mathbb{Z}+1}\int \Dfrtl\epsilon
\frac{\sigma_\gamma(\varepsilon)}{\imag\omega_n - 
\frac{\gamma}{2}(\imag \varphi_m - \Phi)
-\epsilon}
\label{eq:hanansatz}
\end{equation}
for the corresponding self-energy. However, Eq.~\eqref{eq:hanansatz} is only 
approximate, because the $\imag\varphi_m$-dependence of the functions 
$\sigma_\gamma(\epsilon)$ is not taken into account. Such a non-trivial
dependence appears as a result of higher-order corrections in $U$.

Let us start by discussing the analytically continued bare Green's function 
\begin{equation}
G_0(z_\varphi, z_\omega) = 
\sum_{\alpha=\pm 1} 
\frac{\Gamma_\alpha / \Gamma} { z_\omega - \frac{\alpha}{2}(z_\varphi - \Phi) 
+ \imag \Gamma^{(\alpha)}  (z_\phi, z_\omega)},
\end{equation}
with $\Gamma^{(\alpha)} (z_\varphi, z_\omega) := 
\Gamma \sign (\Im z_\omega - \alpha \Im z_\varphi / 2)$.
The corresponding geometric structure of the complex space is depicted in 
Fig.~\ref{fig:cones}, the branch cuts given by the black lines $\gamma=\pm1$.
Note that the Green's function does not vanish for all directions within a
given $T^{C^s_\nu}$ as $\left|\underline z\right| \to \infty$. 
\label{textstelle:G_bounded}
On the other hand, $\Im G_0(\underline z)$ is at least bounded, and we assume that 
nonzero interactions do not alter this fundamental property.
\label{sec:shift_argument}
One can thus always find a constant $c$ such that the imaginary part of the 
function $f(\underline z) :=  G(\underline z) + \imag c$ is positive.
Integral representations of the form $\int f(\zeta) K(z,\zeta) \Dfrtl \zeta =
f(z)$ which are valid for the class of holomorphic functions with non-negative imaginary part also
hold for $G(\underline z)$, since $-\imag c\cdot \mathrm{const}(z)$ is also
a function with non-negative imaginary part.
This class of functions on tubular cone
domains was extensively studied by mathematicians. 
In Ref.~\cite{Vladimirov1978}, Vladimirov finds a generalization of
Herglotz-Nevanlinna representations \cite{Nevanlinna} to such domains. 
See Appendix \ref{sec:AppA} for details. 

The validity of the imaginary-voltage formalism is presently based on the
assumption  of asymptotic convergence of the perturbation series in $U$.
Thus,
the influence of the branch cut between $T^{C^s_{\nu+2}}$ and $T^{C^s_\nu}$ is
expected to become negligible as $\nu\to\infty$, i.e.\ all branch cuts with
$\nu>\nu_\text{crit}$ can be ignored. The maximal value
$\nu_\text{crit}$ may for example be estimated from  
the expansion order histogram of the weak-coupling QMC simulation, since a given branch cut with index 
$\gamma=\nu+1$ is only established by diagrammatic contributions with order larger than a 
certain value $n$, which is roughly proportional to $|\gamma|$.

\begin{figure}
\includegraphics[width=0.49\textwidth]{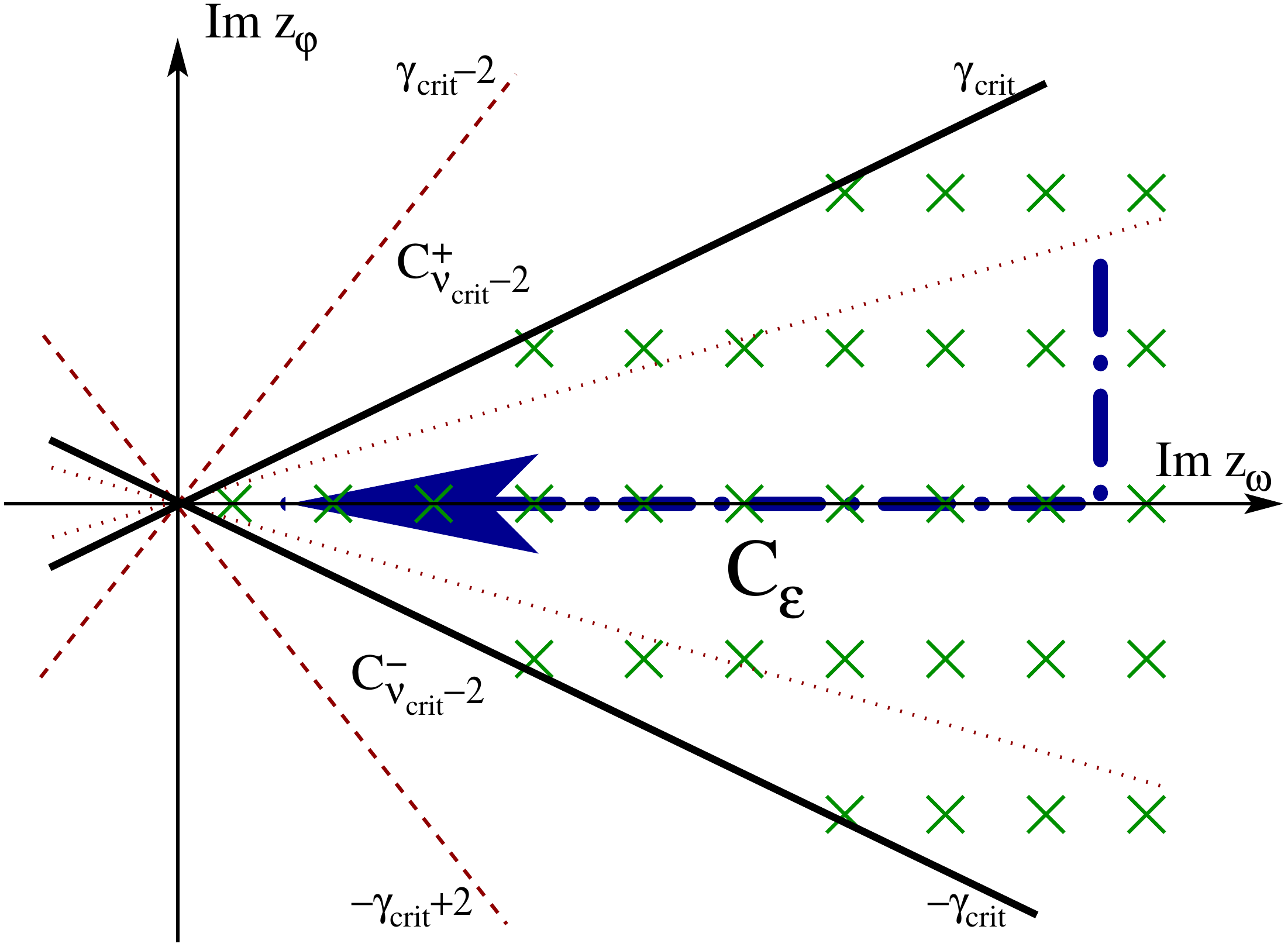}
\caption{(color online)
Sketch of the geometry of the two-dimensional analytic continuation problem. 
For the critical domain index $\nu_\text{crit}$ 
the branch cut $\gamma_\text{crit}+2 = \nu_\text{crit}+1$ (dotted line) 
is negligible, while the critical branch cuts $\pm\gamma_\text{crit}$ 
are not. 
The Green's function is therefore
holomorphic in the cone domain $T^{C_\epsilon}$ bounded by the ratios
$\pm\Im z_\varphi = \frac{2}{\gamma_\text{crit}} \Im z_\omega=:
\epsilon \Im z_\omega$,
with $C_\epsilon$ given by Eq.~\eqref{eq:defepsdomain}. Investigating
the Green's function at the edge of this domain is
compatible with the limiting procedure of taking $z_\varphi\to\Phi$ and
then $z_\omega \to \omega +\imag 0^+$ for the spectral function $A(\omega)$. 
This is indicated by the bold dash-dotted arrow.
Using the integral representation \eqref{eq:kernel_Tceps}, a most 
likely limit of the Green's function at the edge,
$\lim_{\underline \zeta\to \underline x} {G}(\underline z)$, 
$\underline x \in \mathbb{R}^2$,
will be inferred from 
the QMC data $\left.G(\imag\varphi_m,\imag\omega_n)\right|_{T^{C_\epsilon}}$ 
in the domain using a Maximum Entropy Method (Sec.~\ref{MaxEnt}).
The spatial locations of the QMC data points in the domain are 
symbolized by the crosses.
In the case of strong interaction, for small Matsubara frequencies
we are limited to small values of $\varphi_m$.
}
\label{fig:cont2dgeometry}

\end{figure}

As stated in Ref.~\cite{Han07} we are required to first take the limit
$z_\varphi\to \Phi$ and then $z_\omega\to \omega + \imag 0^+$. In our
language, the spectral function is given by 
\begin{equation}
A(\omega) = -\frac{1}{\pi} \lim_{\nu\to\infty}
\lim_{\underline z\to (\Phi, \omega) 
} 
\Im G_{(\nu)}(\underline z).
\end{equation}
Since branch cuts with index $\gamma\geq\nu_\text{crit} + 1$ vanish we choose 
the domain $T^{C_\varepsilon}$ with 
\begin{equation}
C_\varepsilon := \{(x_1,x_2)\in \mathbb{R}^2\,|\, x_2 > 0 \wedge
-\varepsilon x_2 < x_1 < \varepsilon x_2 \},
\label{eq:defepsdomain}
\end{equation}
and $\epsilon \approx \frac{2}{\nu_\text{crit}-1}$ for
the analytic continuation of the interacting Green's function.
This choice of domain is illustrated in Fig.~\ref{fig:cont2dgeometry}.
In practice, the critical branch cut is yet chosen arbitrarily but to be
small, see section \ref{MaxEnt}.
As shown in Appendix \ref{sec:AppA} the Poisson kernel representation resulting from
Vladimirov's theorem is
\begin{equation}
\left.  \Im G(\underline z) \right|_{T^{C_\varepsilon}} 
=
\int_{\mathbb{R}^2}\mathrm{d}^2 x\,
\mathcal{P}_{\epsilon} (\underline z- \underline x) \lim_{\underline \zeta \to \underline x} \left.\Im G(\underline \zeta)\right|_{T^{C_\varepsilon}}
\label{eq:kernel_Tceps}
\end{equation}
with
\begin{equation}
\mathcal{P}_{\epsilon} (\underline z) = 
\frac{1}{\pi^2\epsilon} \prod_{\mu=\pm 1} 
\frac{y_2 - \mu y_1 /\epsilon}
{
(x_2 - \mu x_1 /\epsilon)^2 + (y_2 - \mu y_1 /\epsilon)^2
},
\label{eq:poissonkernelepsdomain}
\end{equation}
where $\underline x$ and $\underline y$ are the real and imaginary parts of
$\underline z$.

\subsection{Maximum Entropy Method}
\label{MaxEnt}

\subsubsection{Single Analytic Continuation}
The numerical analytic continuation of imaginary-time quantum Monte Carlo data is 
a highly ill-posed problem. Even if the finite set of QMC data did not
contain any stochastic noise there 
would exist an infinite-dimensional manifold of solutions to the integral equation
associated with the continuation, i.e.~the spectral representation
\begin{equation}
G(\imag \omega_n) = \int \Dfrtl{\epsilon} \frac{A(\epsilon)}{\imag\omega_n - \epsilon}
=: K_\text{eq}[A] (\omega_n)
\label{eq:kernel1d}
\end{equation}
for the conventional continuation problem.
\par
Hence, a regularization procedure picking a ``most probable" solution is required.
Typically, this is approached with a Maximum Entropy Method
(MEM), a rigorous framework rooted in Bayesian logic
which can be understood as an automatic Ockham's Razor, in the sense of being 
``maximally noncommittal with regard to missing information"
\cite{jarrell_review, maxent_buch, jaynes57}.
The spectral function $A(\omega)$ is interpreted as a probability
distribution.
A default model $D(\omega)$ is introduced as a-priori information about the solution
$A(\omega)$. Additional information, given by the measured imaginary-time data
$\bar G(\imag\omega_n)$, is inferred 
through the kernel $K_\text{eq}[A]$ in \eqref{eq:kernel1d}. If 
there is no additional information the procedure will pick $A(\omega) =
D(\omega)$, in Bryan's MEM algorithm \cite{bryan}.
\par
In practice, a functional 
\begin{equation}
Q[A] = \chi^2[A] - \alpha S[A],\quad \alpha > 0
\end{equation}
is minimized in the space of candidate solutions for a given hyper-parameter
$\alpha$. The QMC data must be Gaussian distributed, such that 
the likelihood penalty $\chi^2[A]$ is given by
\begin{equation}
\chi^2[A] =\frac{1}{2}\sum_{\rho,\eta=1}^N (\bar
G_\rho-K_\text{eq}[A]_\rho)C^{-1}_{\rho \eta}
(\bar G_\eta-K_\text{eq}[A]_\eta),
\label{eq:chisquared}
\end{equation}
where $\bar G_\eta$ are the measured mean real or imaginary parts of the 
imaginary-frequency Green's function $G(\imag\omega_n)$, and  
$C^{-1}_{\rho\eta}$ are the elements of the inverse covariance matrix.
\par
The default model $D(\omega)$ is invoked through
the entropy 
\begin{equation}
S[A] = \int \Dfrtl\epsilon \left[
A(\epsilon) - D(\epsilon) - A(\epsilon) \log \frac{A(\epsilon)}{D(\epsilon)}
\right].
\label{eq:entropy}
\end{equation}
For a detailed theoretical justification of this choice for the entropy see 
Ref.~\cite{maxent_buch}.

The easiest way of fixing the regularization parameter $\alpha$ is to employ 
the condition $\chi^2\approx N$ (historic MEM). It is, however, more reasonable to calculate a
posterior probability distribution $\mathrm{Pr}(A|\alpha)$. 
Setting $\alpha$ to the maximum of the posterior probability distribution is 
called classic MEM. 
Marginalizing $\alpha$ by choosing $\mathrm{Pr}(A|\alpha)$ as weights for
$A$ when integrating over $\alpha$ is empirically found to be most suitable and 
is also most justified from the theoretical point of view (Bryan's MEM).

\subsubsection{Double Analytic Continuation}
In order to adapt the above procedure to the double analytic continuation
problem, a non-negative quantity has to be found which 
\begin{enumerate}
\item uniquely represents any possible function in the data range of
interest -- say $T^{C_\epsilon}$ -- in order to define a $\chi^2$ for
inference;
\item easily allows calculating the non-equilibrium spectral function
$A(\omega)$.
\end{enumerate}
We choose
\begin{equation}
\tilde A(\underline x) := -\frac{1}{\pi} \lim_{\underline \zeta\to\underline x} 
\left.\Im G(\underline \zeta)\right|_{T^{C_\epsilon}}
\end{equation}
as such a representation,
since due to the Kramers-Kronig relations and the validity of the representation
\eqref{eq:kernel_Tceps}, $\tilde A$ yields a unique and simple representation of all
possible functions $G|_{T^{C_\epsilon}}$. The non-equilibrium 
spectral function is easily accessible, since $A(\omega) = \tilde A(\Phi, \omega)$.

In the case of zero interaction,
\begin{equation}
\tilde A_0(\underline x) = -\frac{1}{\pi} \Im\sum_{\alpha=\pm 1}
\frac{\Gamma_\alpha/\Gamma}
{
x_2 - \alpha (x_1 - \Phi) / 2 - \epsilon_d + \imag \Gamma
}.
\end{equation}
It is easy to verify that $\tilde A_0(\underline x)$ is a positive
function with $\int\mathrm{d}^2x \tilde A(\underline x)=l$ if one constrains the
$x_1$-integration to an arbitrary finite interval of length $l$. 
This fact and the fact that $\tilde A(\Phi,\omega) = A(\omega) \geq 0$ do
not imply $\tilde A(x_1,x_2)\geq 0$ in general.
We however assume $\tilde A(x_1,x_2) \geq 0$ and expect to obtain 
revealing signatures within the MEM, 
in case the real $\tilde A$ is not positive definite for a
given data set. Note that even in the presence of regions where $\tilde{A}<0$, 
a MEM can be implemented, by identifying the nodes of $\tilde A$,
as in the case of bosonic spectral functions.
In general, positivity may be enforced by adding a positive real constant $b$
to the spectral function and adding a corresponding term to the image.
As particular example for this procedure, we quote here the
case of the Nambu off-diagonal Green's function $G_{12}$, where the
positivity is enforced as $G_{12}(\tau) + b \int \Dfrtl\omega K(\tau,\omega) =
\int \Dfrtl\omega K(\tau,\omega)(A_{12}(\omega)+b)$ \cite{jarrell}.

We hence choose \eqref{eq:kernel_Tceps} as a kernel function for the
$\chi^2$ functional and only take data in $T^{C_\epsilon}$ into
account. 
 The entropy expression \eqref{eq:entropy} is adopted for a 
two-dimensional default model $\tilde D(\underline x)$. 

\subsubsection{Implementation}
First note that since the input data for the Poisson kernel 
\eqref{eq:kernel_Tceps} are obtained from statistically independent QMC simulations, the
covariance $C$ in the $\chi^2$ functional \eqref{eq:chisquared} has a 
block-diagonal shape
\begin{equation}
C = 
\begin{pmatrix}
C^{(m_\text{min})} &               0      & \cdots  &          \\
   0               & C^{(m_\text{min}+1)} &    0    & \cdots   \\ 
   \vdots          &  0                   &  \ddots    &    \\ 
                   &   \cdots             &  0    &   C^{(m_\text{max})} \\ 
\end{pmatrix}.
\label{eq:covariance}
\end{equation}
The submatrices $C^{(m)}$ are covariances for the subset of data
$G(\imag \varphi_m,\imag\omega_n)$ at a fixed $\varphi_m$, estimated from
the output of the corresponding equilibrium QMC simulation.
\par
Our implementation of the Maximum Entropy Method is based on Bryan's standard 
algorithm introduced in Ref.~\cite{bryan}. 
A singular value decomposition (SVD) of the kernel
\begin{equation}
K : V_{\tilde A} \to V_\text{data}, \quad
K = V \Sigma U^T
\label{eq:singvaldecomposition}
\end{equation}
is performed, with $V$, $U^T$ orthogonal, and the singular values
\begin{equation}
\Sigma = \mathrm{diag}(\sigma_1, \sigma_2, \dots, \sigma_s, 0, \dots, 0),
\end{equation}
$\sigma_1 \geq \sigma_2 \geq \dots \geq \sigma_s > 0$.
  Many important quantities may be reduced to the $s$-dimensional
singular space $V_\Sigma$. 
Most notably, the $(\dim V_{\tilde A})$-dimensional optimization problem given by 
\begin{equation}
Q[\tilde A]\stackrel{!}{=}\text{min}
\label{eq:optimprob}
\end{equation}
may be solved within the singular space using Levenberg-Marquardt iterations. 
As $s$ is comparably small
after truncating the singular space with respect to the floating point precision 
of the singular values $\sigma_i$
(typically, $s \approx 50$), the algorithm is still sufficiently efficient, 
even though a two-dimensional frequency grid is required for 
the numerical resolution of $\tilde A$, and hence 
$\mathrm{dim}\,V_{\tilde A} \geq 10^5$.
\par
The algorithm enables us to calculate several important data qualifiers and 
posterior probabilities and therefore to classify both input data quality and 
candidate solutions.
The posterior
\begin{equation} 
\mathrm{Pr}(\alpha|\bar G)
= \mathrm{Pr}(\alpha)\int \mathcal{D} \tilde A\, \frac{\euler{Q}} {Z_L Z_S(\alpha)},
\end{equation}
with $Z_L = \int \mathcal{D} [K\tilde A]\, \euler{-\chi^2/2}$, 
$Z_S(\alpha) = \int\mathcal{D} \tilde A\, \euler{\alpha S}$,
and the Jeffreys prior \cite{Jeffreys46} $\mathrm{Pr}(\alpha) \propto \alpha^{-1}$, is calculated using
a Gaussian approximation for $Q$, centered around the solution
$\tilde A_{\text{opt}, \alpha}$ of Eq.~\eqref{eq:optimprob}.
\par
The usual procedures and strategies for data qualification and improvement of
results as described in \cite{jarrell_review} are adopted:
Assuming a flat prior $\mathrm{Pr}(\tilde D)$, the posterior for the default
model
\begin{equation}
\mathrm{Pr}(\tilde D|\bar G) \propto
\int \Dfrtl \alpha \int\mathcal{D}\tilde A\,
\mathrm{Pr}(\alpha) \frac{\euler{Q}}{Z_L Z_S(\alpha)}
\end{equation}
is computed easily. 
$\mathrm{Pr}(\tilde D|\bar G)$ serves as evidence for the quality of prior
information when comparing within sets of default models for given QMC data.
Whereas a posterior probability for the domain parameter $\epsilon$ for given
data and given default model, $\mathrm{Pr}(\epsilon|\bar G,\tilde D)$, would be a 
sensible extension to the algorithm, we have not derived it yet.
Useful ingredients might be found in the literature on blind deconvolution in 
signal processing, see \cite{pinchas05}.
In our implementation, a small enough $\epsilon$ is chosen a priori.
\par
Picking appropriate data sets with well-estimated covariance from the QMC 
output is also a non-trivial part of the problem. 
A good check is to determine the most probable mock error rescaling $\sigma$
where 
the covariance $C$ is formally substituted by $\sigma^2 C$. If the most
probable $\sigma$ (``merit"), i.e. the solution of
\begin{equation}
\frac{\chi^2_\text{classic}}{\sigma^2} + N_g = N
\end{equation}
deviates from 1 by more than a few tens of percent, the input 
data are rejected \cite{jarrell_review}. 
$\chi^2_\text{classic}$ is the $\chi^2$ value of the classic MEM solution, 
the number of data points $N$, and the number of ``good" data points
$N_g = \sum_i \frac{\lambda_i}{\alpha_\text{classic} + \lambda_i}$ with $\lambda_i$ the
eigenvalues of 
\begin{equation}
\Lambda_{ij} = \left[\sqrt{\tilde A_i} \frac{\partial^2
\chi^2/2}{\partial
\tilde A_i \partial \tilde A_j} \sqrt {\tilde A_j}\right]_{\tilde
A_\text{classic}}.
\end{equation}
In practice, a maximal Matsubara
frequency $n_\text{max}$ compatible with the error rescaling merit was 
determined, and all data $\Im G(\imag \varphi_m, \imag \omega_n)$ in 
$T^{C_\epsilon}$, with $n \leq n_\text{max}$ were used for inference.
Presumably, better data selection strategies do exist.
For example, using independent measurements for $\Re G$ and
taking them into account by using a Schwarz representation (see
Appendix \ref{sec:AppA}) could yield better results. Furthermore, 
the largest Matsubara frequency index $n_\text{max}$ could be determined 
for each $\varphi_m$ individually. The latter appears to be necessary for
non-equilibrium data.
\par
For the truncation of singular values, a threshold $\lambda$ was used,
\begin{equation}
\sigma_i \mapsto 
\begin{cases}
\sigma_i, & \text{if } \sigma_i\geq \lambda \sigma_1 \max \{M,N\}, \\
0 , & \text{else}
\end{cases}
\label{eq:singvalthreshold}
\end{equation}
for an $M$ by $N$ kernel matrix.
While for the conventional Wick rotation $\lambda\approx 10^{-8}$ was
sufficient, $\lambda \approx 10^{-12}$ had to be chosen in our case in order to take all
relevant search directions in the $\tilde A$ space into account. Quadruple precision
floating point arithmetic was found to be unnecessary. For discretizing the $\tilde A
(\underline x)$ function, logarithmic meshes 
for the $x_1$ and $x_2$ variables
were used. Although $\tilde A(\underline x)$ does not decay for
all directions as $\underline x\to \infty$, choosing a finite mesh and 
truncating the integrals was not found to be critical.

\subsubsection{Equilibrium}

\begin{figure}
\includegraphics[width=0.49\textwidth]{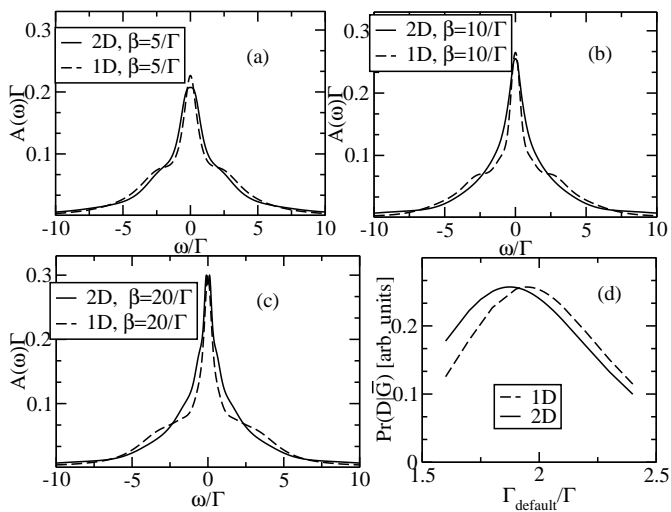}
\caption{Analytically continued data for equilibrium ($\Phi=0$) obtained using the conventional
Wick rotation (1D) and the unconventional two-variable continuation (2D), with
$U=5\Gamma, V_G=0$. The domain parameter for 2D continuation is 
$\varepsilon = \frac{2}{19}$.
Subfigure (d) shows the posterior probabilities $\mathrm{Pr}(D|\bar G)$
of the default models as a function of $\Gamma_\text{default}$ (Eqs.~\eqref{eq:def_eq_1d} and \eqref{eq:def_eq_2d}).
}
\label{fig:spectra_equilibrium}
\end{figure}

As a test case we consider the equilibrium limit $\Phi=0$. The data for 
$\varphi_m=0$ can be analytically
continued with the standard Wick rotation, using Eq.~\eqref{eq:kernel1d} and
the standard MEM. 
Figure~\ref{fig:spectra_equilibrium} compares this 1D spectral function to the result
based on the 2D data set  
$G(\imag\varphi_m,\imag\omega_n)$ and continued using the domain 
$T^{C_\varepsilon}$ and the kernel function defined in Eq.~\eqref{eq:kernel_Tceps}.
As default models for high temperatures we use Lorentzians with variable width
$\Gamma_\text{default}$.
They read
\begin{equation}
D(\omega) = \frac{1}{\pi}
\frac{\Gamma_\text{default}}{\omega^2 + \Gamma_\text{default}^2}
\label{eq:def_eq_1d}
\end{equation}
for the 1D continuation and
\begin{equation}
\tilde D(x,\omega) = 
\frac{1}{\pi} \frac{\tilde\Gamma_\text{default}(x)}
  {
    \omega^2 + \tilde\Gamma_\text{default}(x)^2
  }
\label{eq:def_eq_2d}
\end{equation}
for the 2D continuation,
with $\tilde\Gamma_\text{default}(x) = \sqrt{\Gamma_\text{default}^2 + x^2}$.
An annealing procedure in the temperature was used for both, the 1D and 2D data
for invoking adequate prior information, i.e.~we used the 
$\tilde A$ solution of the next higher temperature as default model, starting
with the Lorentzian at the highest temperature. 
This default model selection procedure appears not to have any strict Bayesian 
justification, however the physical argument is freezing out the high-frequency 
degrees of freedom and using present data for inferring low-energy details of the 
spectrum step by step \cite{jarrell}. 
A similar idea plays the key role in several modern renormalization group 
techniques.
Note that Gaussian default models are not well-suited for our data, since
the high-frequency tail in the wide-band limit is Lorentzian. 
This manifests itself quantitively in the following way: For the Gaussian default 
models we tested all had $\mathrm{Pr}(\tilde D|\bar G)$ one order of magnitude 
lower than the Lorentzian ones.
For both, the Gaussian and the Lorentzian, we can expect the quantity 
$\Gamma_\text{default}/\Gamma$ to be $>1$, due to the overall broadening 
introduced by a finite interaction $U$.

Indeed, for the parameters $U=5\Gamma, V_G=0$, $\Phi=0$
shown in Fig.~\ref{fig:spectra_equilibrium}(d), the
(unnormalized) posterior probabilities $\mathrm{Pr}(D|\bar G)$ and
$\mathrm{Pr}(\tilde D|\bar G)$ 
as a function of the parameter $\Gamma_\text{default}$ are peaked at $\approx
2\Gamma$ for both, the 1D and 2D continuation procedures, respectively.
These probabilities were calculated for $\beta\Gamma=10\Gamma$.  The most probable
$\Gamma_\text{default}$ was chosen as default model. However, a strong dependence
of the results on $\Gamma_\text{eff}$ was not observed.

The spectral functions shown in Fig.~\ref{fig:spectra_equilibrium} were obtained for $\beta\Gamma=5$, $10$, and $20$.
We chose $\varepsilon = \frac{2}{19}$ for the 2D domain, using
$n_\text{max} = 10$, $20$, $40$ for $\beta\Gamma=5$, $10$, and $20$, respectively.
Note that due to the simple data selection strategy described in the previous section we 
only took into account data points with $\varphi_{-2}\leq\varphi_m\leq
\varphi_2$. 
Using a global $n_\text{max}$, the estimate for the covariance 
submatrix $C^{(m=0)}$ in Eq.~\eqref{eq:covariance} eventually becomes singular, even though $C^{(m)}$
with $|m|\geq 3$ and $\omega_n>\omega_{n_\text{max}}$ could still be estimated for a limited set of Matsubara
frequencies. 
We expect that using such additional,  
well-estimated $C^{(m)}$  
might lead to more structured spectral functions. 
In practice, however, the merit $\sigma$ must
yet be viewed as a rather crude measure of the quality of the covariance estimate. So for
the purpose of both simplicity and reproducibility we used the stronger restriction.

The Kondo temperature for 
$U=5\Gamma$ 
is $T_{\rm K}/\Gamma\approx 0.1$, i.e.\
we can expect first signatures of strong coupling physics like Hubbard bands and
a temperature dependent quasi-particle peak of reduced width in the spectra.
Indeed both the 1D and 2D MEM reproduces these features. More importantly,
the overall shape of the spectra obtained agrees for all temperatures 
shown in Fig.~\ref{fig:spectra_equilibrium}(a-c).
The results depend only slightly on the choice of $\Gamma_\text{default}$
for relevant values of $\mathrm{Pr}(\tilde D|\bar G)$.
Although the spectra inferred from the 2D procedure using our current implementation appear to be less structured, the overall shape seems to
be reconstructed quite well. For more serious calculations, the detailed 
high-frequency behavior (and behavior for large $x$)
should be introduced with 
a more sophisticated default model, e.g.\ based on perturbation theory.

\subsubsection{Inferred Representation}
\begin{figure}
\includegraphics[width=0.49\textwidth]{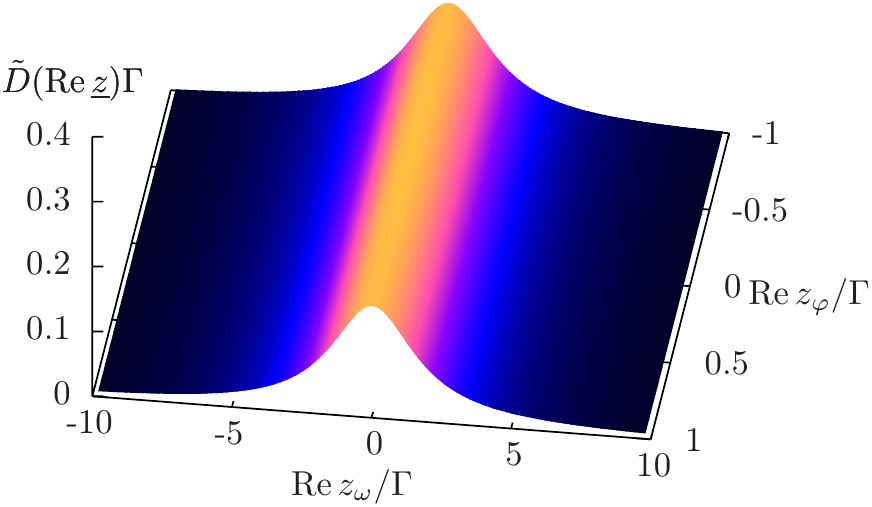}
\caption{(color online) Lorentzian default model \eqref{eq:def_eq_2d} with best
$\mathrm{Pr}(\tilde D|\bar G)$ for first annealing step in equilibrium.}
\label{fig:defaultAtilde}
\end{figure}

\begin{figure}
\includegraphics[width=0.49\textwidth]{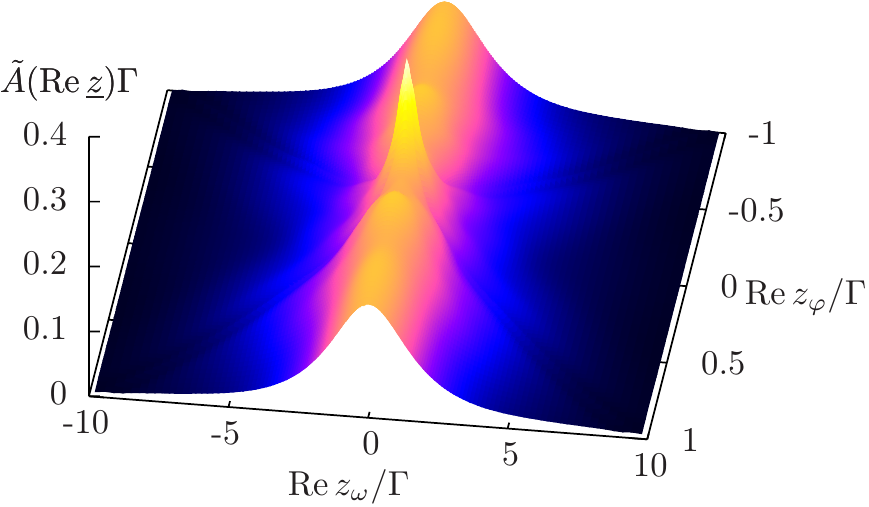}
\caption{(color online) MEM solution for $\tilde A$ 
inferred from the QMC data at
the lowest temperature, $\beta\Gamma=20$, for the equilibrium test case shown
in Fig.~\ref{fig:spectra_equilibrium}.}
\label{fig:inferredAtilde}
\end{figure}

Figure~\ref{fig:defaultAtilde} shows the Lorentzian default model we used at
the highest temperature in the annealing procedure, $\beta\Gamma=2$.
At the lowest temperature $\beta\Gamma=20$, the representation shown in Fig.~\ref{fig:inferredAtilde} was obtained. The equilibrium spectral function
shown in Fig.~\ref{fig:spectra_equilibrium}(c) is given by the cut $\tilde
A(\Phi = 0,\omega)$. Other values of $\Re z_\varphi$ do not have any physical
meaning.
Note that certain structures appear in the inferred $\tilde A(x,y)$ which
vary as the domain parameter $\epsilon$ is changed: they occur for 
$\Re z_\varphi = \pm \Re \epsilon z_\omega$. 
We interpret them as resulting 
from the 
properties of the kernel function discussed in Sec.~\ref{sec:kernelstructure} in combination with the MEM principle of only
incorporating changes which are strongly supported by data.
Also, at larger distance from the origin, discretization errors from the
discretization of the double integral are most dominant for this most structured
region of the kernel.
At finite bias, the qualitative structure of the inferred representation 
remains unchanged. 

\subsubsection{Finite Bias}

The rule of thumb  $n_\text{max}\approx\frac{\beta U}{2}$ appeared to be a good 
choice for preparing the equilibrium QMC data for inference. 
For $\Phi>0$ a first interesting observation is that at sufficiently low 
temperatures $n_\text{max}$ seems to be considerably smaller than $\frac{\beta U}{2}$.
\par
In fact, the simple data selection strategy yielding $n_\text{max}$ does
not appear to produce a sufficiently informative data set to obtain quantitative 
agreement with for example RT-MC calculations \cite{Werner10}.
We observed this problem for $\beta\Gamma=10$ and the interaction strengths 
$U=4\Gamma$ and $U=6\Gamma$ and several values of the bias voltage $\Phi$.
On the other hand, by picking an $n_\text{max}$ for each $\varphi_m$ separately, we 
found larger sets of admissible input data, which tend to
show a good agreement with RT-MC data for the
current-voltage characteristics. While the procedure is yet somewhat arbitrary,
the following criteria were used to restrict the choices of data
sets producing convergent MEM solutions:
\begin{itemize}
\item ensure an error rescaling $\sigma\approx 1$;
\item discard strongly oscillating solutions 
and 
solutions with obvious artifacts around $\omega\approx 0$;
\item discard solutions which strongly violate the physical
sum rule $\|A\|:=\int \Dfrtl\omega A(\omega) = 1$. In
many cases, too small values $\|A\|\approx 0.9$ were obtained. Note that the
MEM as we implemented it only has prior information about the value of the 
truncated double integral 
$\iint\mathrm{d}^2x \tilde A(\underline x)$, because two-dimensional
probability densities are considered when the entropy expression 
\eqref{eq:entropy} is straightforwardly generalized with respect to $\tilde
A$;
\item use as many data points as possible, starting with small $\omega_n$,
to maximize the amount of accessible information. 
\end{itemize}
Note that the domain parameter $\epsilon$ was, again, chosen somewhat arbitrarily:
For $U=4\Gamma$ we only investigated $\nu_\text{max}=16$, for
$U=6\Gamma$ we picked $\nu_\text{max}=20$, with 
$\epsilon = \frac{2}{\nu_\text{max}-1}$.
The dependence of the results on the particular choice of $\epsilon$ was not 
studied systematically yet, but work along these lines is under way and
the results will be presented elsewhere.
The usual annealing procedure with temperatures $\beta\Gamma=2$,
$\beta\Gamma=5$, $\beta\Gamma=10$, where for $\beta\Gamma=2$ the Lorentzian
default models with $\Gamma_\text{default}=1.5\Gamma$ ($U=4\Gamma$) and 
$\Gamma_\text{default}=2.1\Gamma$ ($U=6\Gamma$) were found to be most
suitable based on the posterior $\Pr(\tilde D| \bar G)$.
\par
The current $J$ was computed using Meir and Wingreen's equation 
\cite{meirwingreen94}
\begin{equation}
J = J_\text{max} \int \Dfrtl \omega
\left[f_L(\omega)-f_R(\omega)\right] A(\omega),
\label{eq:meirwingreen}
\end{equation}
with $J_\text{max} = \frac{\Gamma e}{h}$.

Our experience up to now indicates that for too small sets of QMC data the 
method systematically underestimates the current, because Bryan's algorithm 
by convention does not incorporate any changes to
$\tilde A \approx \tilde D$ in case the data do not provide sufficient 
evidence for such modifications. As a result, the current is too small, because 
in the vicinity of $\omega\approx 0$
the less structured
default model obtained from the next higher temperature (initially the broad
Lorentzian \eqref{eq:def_eq_2d}) is much flatter than the
true solution, which features a sharp Abrikosov-Suhl resonance in the relevant
frequency range. Hence, the spectral function obtained from the MEM has less spectral 
weight in the integration window in Eq.~\eqref{eq:meirwingreen}
than the true $A(\omega)$.
\begin{figure}
\includegraphics[width=0.49\textwidth]{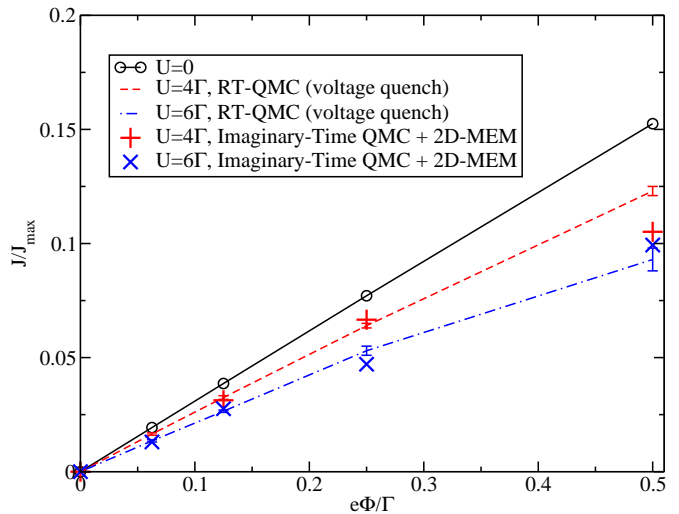}
\caption{(color online) Current-voltage characteristics obtained using the 2D MEM compared to RT-QMC
\cite{Werner10} data for indicated Coulomb interactions at temperature $\beta\Gamma=10$.
}
\label{fig:vgl_rtqmc}
\end{figure}

Due to this trend of underestimation, in Fig.~\ref{fig:vgl_rtqmc} we compare the 
largest values of the current compatible with the above-listed restrictions
to data obtained using a recently developed RT-MC approach \cite{Werner10}.
A generally good agreement is obtained. However, the data selection procedure is still
too arbitrary to consider these results unbiased.
Error bars are not available. If we only considered 
a fixed set of data $\bar G$, the
covariance $\text{Cov}(\tilde A(\underline x^{(1)}), \tilde A(\underline
x^{(2)}))$ would be estimated easily \cite{bryan}. However, due to large off-diagonal
terms, attempting to estimate an error bar for $J$ is rather cumbersome. 
The $\Phi/\Gamma=0.0625$ run did not converge to a solution meeting our
criteria for $U=4\Gamma$. 

\subsubsection{Non-Equilibrium Spectral Functions}
\begin{figure}
\includegraphics[width=0.49\textwidth]{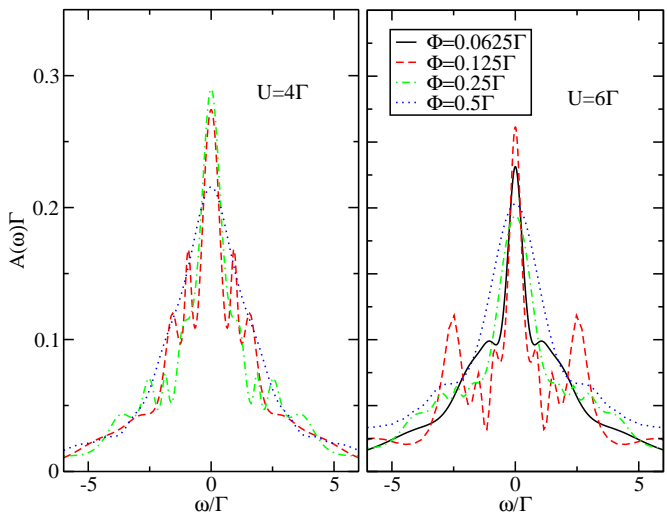}
\caption{(color online) Spectra $A(\omega) = \tilde A(\Phi,\omega)$ used for the computation of the current shown in 
Fig.~\ref{fig:vgl_rtqmc}. 
}
\label{fig:spectranoneq}
\end{figure}

\begin{figure}[t]
\includegraphics[width=0.49\textwidth]{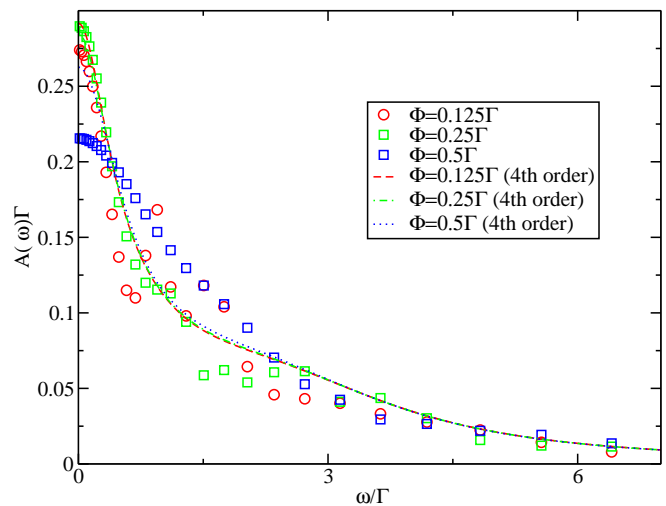}
\caption{(color online) Spectra $A(\omega) = \tilde A(\Phi,\omega)$ for $U=4\Gamma$ as
compared to fourth-order perturbation theory.
}
\label{fig:spectrapth}
\end{figure}

Spectra resulting from the procedure described above are shown 
in Fig.~\ref{fig:spectranoneq}. These are the spectral functions used to compute the current in Fig.~\ref{fig:vgl_rtqmc}.
While oscillations appear, presumably due to the neglected error of the 
covariance estimate \cite{vonderlinden},  
it is evident that the 
overall spectral weight at
small $\omega$ is larger for $U=4\Gamma$ than for $U=6\Gamma$ when $\Phi <
0.5\Gamma$.
This is consistent with the expectation that the quasi-particle resonance
for $U=6\Gamma$ is already suppressed, because $\beta^{-1}=0.1 > T_K$, whereas 
$\beta^{-1} \approx T_K$ for $U=4\Gamma$.
In Fig.~\ref{fig:spectrapth}, we show a comparison of the spectral functions for $U/
\Gamma=4$ and $\beta\Gamma=10$ to the result obtained from fourth-order 
perturbation theory \cite{fujiiueda03}. Based on the results presented in Ref.~\onlinecite{Werner10}, we expect that fourth-order perturbation theory is quite accurate at this interaction strength and temperature.  Besides 
the unphysical oscillations in the MEM result and a bias towards the high-temperature default model, especially for larger voltage biases, the agreement 
between the spectral functions, in particular the qualitative distribution of
the spectral weight, seems satisfactory.

\par
Table \ref{tab:dataselU4} shows the norm $\|A\| = \int A(\omega) \Dfrtl\omega$
for the functions presented in the figure. 
\begin{table}
\begin{center}
\begin{tabular}{|c||c|c|}
$\Phi/\Gamma$ & $\|A\|_{U=4\Gamma}$ & $\|A\|_{U=6\Gamma}$ 
\\
\hline
0.0625        &   --                  &  0.91  \\
0.125         &   0.92                &  0.92  \\
0.25          &   0.92                &  0.95  \\
0.5           &   1.03                &  1.16  
\end{tabular}
\end{center}
\caption{Norms of the spectral functions shown in Fig.~\ref{fig:spectranoneq}.}
\label{tab:dataselU4}
\end{table}
Obviously, the physical sum rule $\|A\|=1$ is not strictly obeyed, and there
is a slight tendency towards too small norms whose origin is unclear but
which appears to be consistent with the trend of current underestimation.
Moreover, the selection of data we chose at
$\beta\Gamma=10$ for $U=4\Gamma$ and $U=6\Gamma$ is shown
in table \ref{tab:dataselU4} and table \ref{tab:dataselU6}, respectively.
The tables present the number $N_m\approx n_\text{max}(m)-2 m/\varepsilon$ of Matsubara frequencies 
which are located within the cone domain $T^{C_\varepsilon}$ for the chosen
$n_\text{max}(m)$.
\begin{table}
\begin{center}
\begin{tabular}{|c||c|c|c|c|}
$\Phi/\Gamma$ & $N_{m=0}$ & $N_{m=\pm1}$ & $N_{m=\pm2}$ & $N_{m=\pm3}$
\\
\hline
0.0625        &   --  &  --  &  -- & --  \\
0.125         &   26  &  12  &   6 &  3  \\
0.25          &   24  &  12  &   5 &  3  \\
0.5           &   24  &  12  &   6 &  3  
\end{tabular}
\end{center}
\caption{
Number $N_m$ of Matsubara frequencies taken into
account for each value of $m$ taken into account in
the data selection at $\beta\Gamma=10$ for $U/\Gamma=4$ and 
for the voltages plotted in Fig.~\ref{fig:vgl_rtqmc}.
}
\label{tab:dataselU4_double}
\end{table}
\begin{table}
\begin{center}
\begin{tabular}{|c||c|c|c|c|}
$\Phi/\Gamma$ & $N_{m=0}$ & $N_{m=\pm1}$ & $N_{m=\pm2}$ & $N_{m=\pm3}$
\\
\hline
0.0625        &   20  &  11  &   6 &  1  \\
0.125         &   21  &  11  &   6  & 8   \\
0.25          &   21  &  11  &   6 &  3  \\
0.5           &   20  &  11  &   6 &  1  
\end{tabular}
\end{center}
\caption{
Same as Table \ref{tab:dataselU4} but for $U/\Gamma = 6$.}
\label{tab:dataselU6}
\end{table}
We did not consider larger values of $m$, although
at least $m=\pm 4$ yields further relevant information about $\tilde A$.
For a test case the spectra did not show dramatic qualitative changes as 
additional values at larger $\varphi_m$ were included, as long as the error
scaling merit remained $\sigma\approx 1$. However, the level of
arbitrariness in the data selection would have been even larger, because of the
corresponding additional $n_\text{max}$ parameters.

Obtaining reliable spectral functions at finite bias will obviously require more effort
and we will briefly comment on possible avenues for this effort in the Conclusion.

\subsubsection{Kernel Structure}
\label{sec:kernelstructure}
We finish with some remarks about the structure of the kernel function 
\eqref{eq:kernel_Tceps} and its role in the continuation problem.
In the language of Bayesian inference the kernel function defines the 
information channel through which evidence about the shape of the 
representation function $\tilde A(\underline x)$ and thus also the 
physical spectral function $A(\omega)$ is extracted from the Monte Carlo 
data.

For the information provided by a single data point, this 
channel results in vague (strong) evidence for changes 
in a given compact region $R\subset V_{\tilde A}$, see 
Eq.~\eqref{eq:singvaldecomposition}, depending on whether the
subset of column vectors $u_i$ of $U$ spanning $R$ is associated with 
small (large) singular values $\sigma_i$, and a small 
(large) overlap of the column vectors $v_i$ of $V$ with the data 
point.
For this reason, very small singular values yield irrelevant components of
the channel and are therefore projected out in Bryan's algorithm by
introducing the threshold $\lambda$, Eq.~\eqref{eq:singvalthreshold}.
\par
We can neither perform the SVD analytically, nor can we analytically take 
into account structural changes which occur when rotating the basis of 
$V_\text{data}$ to the eigenbasis of the covariance matrix $C$ in order to 
consider statistically independent data. 
We can however consider values of the kernel in $ V_{\tilde A}$ for a 
given data point,  assuming it to be uncorrelated with other data points
so that it may be investigated separately. Within our QMC implementation, 
experience shows that correlations
between Matsubara frequencies $\omega_{n}$, $\omega_{n'}$ are monotonically 
decreasing as a function of distance $|\omega_{n'}-\omega_{n}|$, though very 
slowly.

Let us first consider a single uncorrelated imaginary part of a Green's
function at Matsubara frequency $\omega_n$ in the standard Wick rotation 
problem. 
The spectral function $A(\epsilon)$ is inferred through the Lorentzian-shaped
kernel \eqref{eq:kernel1d},
\begin{equation}
\Im K_\text{eq}[A(\epsilon)](\omega_n)
 =
- \frac{\omega_n}{\epsilon^2 + \omega_n^2}.
\end{equation}
For all $\omega_n$ the kernel \eqref{eq:kernel1d} is centered around $\epsilon=0$ and higher frequencies are
associated to larger values of the kernel as the width given by $\omega_n$ is
increased. As compared to $\epsilon\approx 0$ the values of the kernel
at large frequencies are still small.
We can therefore expect large singular values and thus relevant
components of the kernel to be associated with small frequencies only.
This is in agreement with the well-known observation that high-frequency 
information about the spectral function is better put into the default model 
as prior knowledge and a good resolution is obtained for the -- fortunately
most interesting -- low-frequency region.

\begin{figure}
\includegraphics[width=0.49\textwidth]{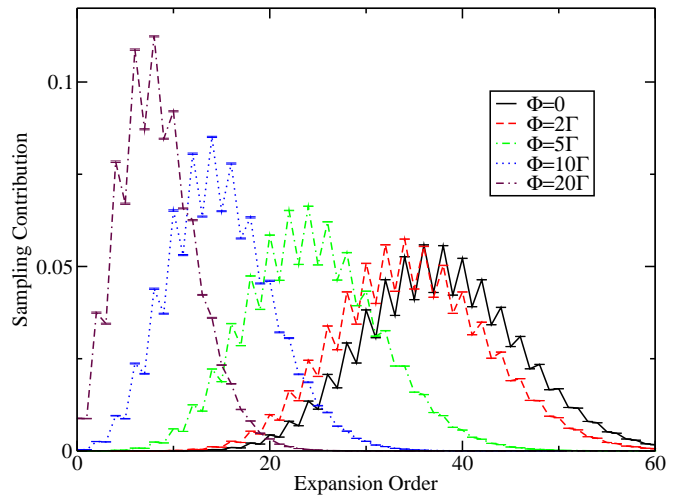}
\caption{(color online) Expansion order histogram obtained using the weak-coupling solver
for $K=-\beta U/4 + 1$, which suppresses the odd perturbation orders \cite{Werner10}.
The results are for $U=7\Gamma$, $\beta\Gamma=51.2$, and 
$\varphi_m=2.46\Gamma$ for indicated voltages $\Phi$. 
The average order decreases as $\Phi$ is increased.
A similar behavior is obtained for different values of $\varphi_m$ and $U$.
}
\label{fig:exporderhist}
\end{figure}

In the case of our two-dimensional continuation the situation is quite similar.
For given data $\Im G(\imag \varphi_m, \imag\omega_n)$ the Poisson kernel 
in Eq.~\eqref{eq:kernel_Tceps} is 
$$
\frac{1}{\pi^2 \epsilon}
\prod_{\mu=\pm 1}
\frac{
  \omega_n - \mu\varphi_m/\varepsilon
}
{
  (x_2 - \mu x_1 / \varepsilon)^2 + (\omega_n - \mu \varphi_m /
\varepsilon)^2
}.
$$
It is the product of two Lorentzians. In analogy to the argument given above
one may expect the best resolution for data $\tilde A(\underline
x^\text{(best)})$ with 
\begin{equation}
x_2^\text{(best)} \approx \pm x_1^\text{(best)} / \varepsilon\ \text{and }
x_2^\text{(best)}, x_1^\text{(best)} \approx 0.
\label{eq:bestresolved}
\end{equation}
This does not depend on the physical voltage $\Phi$, except that the critical
branch cut index $\gamma_\text{crit}$ appears to be decreasing as a function 
of $\Phi$.
This can be estimated from the expansion order histogram for the example
shown in Fig. \ref{fig:exporderhist}.
Consequently, the domain parameter $\varepsilon$ could presumably be raised as 
$\Phi$ be increased. However, in the limit of very large voltages, 
especially the low-frequency region of the physical spectrum 
$A(\omega) = \tilde A(\Phi,\omega)$  
is not expected to be in the best resolvable region \eqref{eq:bestresolved}.
\par
Thus, the approach based on a representation of data in $T^{C_\epsilon}$ 
appears to be limited to relatively small voltages. Note that, since
$\Phi\approx T_K$ is the most interesting parameter regime, this is
presumably no serious drawback. However, identifying subtle details in the range 
$-\Phi/\varepsilon \ll \omega \ll \Phi/\varepsilon$ may require more care
than the case $\omega\approx 0$ for the standard Wick rotation.
Fixing the $x_2$ and $x_1$ 
variables
in the kernel and analyzing the dependence as a function of the
data coordinates $\varphi_m$, $\omega_n$ we similarly find that large values of
 the kernel are found in the vicinity of the domain boundary, i.e. for
$(m,n)$ pairs close to the cone boundary,
$\omega_n\approx \pm \varphi_m/\varepsilon$, with $\varphi_m$, $\omega_n$
not being too large. Hence, data close 
to the boundary provide the most relevant information. This appears to explain 
the importance of an $m$-dependent $n_\text{max}$ in our computation of
non-equilibrium spectra.

\section{Conclusion and Perspective}
\label{Conclusion}
The imaginary time formulation for steady state 
transport in strongly correlated quantum impurity systems proposed by Han and Heary
is based on the solution of a family of quantum impurity models subject to complex voltages, and a subsequent double 
analytical continuation with respect to frequency and voltage.
A main purpose of the study presented in this paper was to investigate to what 
extent an unbiased, numerical 
implementation of this approach is feasible and whether or not it yields 
physically plausible results.

To solve the impurity problem we employed two 
recently developed continuous-time impurity solvers. The hybridization 
expansion approach was found to be unsuitable in the case of large complex 
voltages, due to a serious sign problem resulting from the shift of the 
hybridization function to negative values. The weak-coupling approach, on the 
other hand, 
works well for small and large 
$\varphi_m$. 
Even though the non-interacting Green's function $G_0$ becomes 
complex and oscillating, the resulting sign problem is mild, enabling us to 
obtain highly accurate, unbiased imaginary-frequency data for all relevant 
complex voltages. This part of the problem can be considered as solved, leaving us with the double analytical continuation problem. 

A main result of this work is the derivation of an analytical expression of the
kernel (Eq.~(\ref{eq:poissonkernelepsdomain})) for the analytical continuation 
procedure. 
This kernel is consistent with
the analytical structure (branch cuts) of the theory and maps a function of two
variables, $\tilde A(x_1, x_2)$, to the interacting Green's function in a tubular
cone domain of the complex voltage and frequency space. The physical spectral 
function for a dot under voltage bias $\Phi$ is obtained as $A(\omega)=\tilde A(\Phi,\omega)$. 

We have implemented and tested an analytical continuation procedure
based on the Maximum Entropy Method and our proposed kernel. 
We want to emphasize that both the data selection procedure and the
estimate of the covariance entering into the maximum entropy
employed for $\Phi>0$ are at this point still rather rudimentary and leave room for improvement. 
Our results for the non-equilibrium case should therefore 
be viewed as preliminary and illustrate the presently most plausible spectral functions and currents 
which can be obtained using our current implementation. 

Nevertheless, taking into account the obvious challenges inherent in a double analytical continuation procedure, we
find physically reasonable spectral functions for the interacting 
equilibrium model and, to a lesser extent, also under finite bias. 
A comparison of the spectral functions with fourth-order perturbation theory
shows that the approach is able to reproduce the correct trends,
albeit the strong oscillations resulting from the maximum entropy approach render a detailed
comparison meaningless.
On the other hand, the current calculated using these spectral functions is in fair agreement
with recent results from a real-time Monte-Carlo approach. 

We hope that further improvements in data selection strategies, a better understanding of 
the precise  behavior of the Green's function across the branch cuts, 
improved default model functions and, very importantly, the inclusion of the sum rules
into the maximum entropy algorithm
will eventually enable us to obtain more accurate results and turn the
combination of Monte-Carlo and double analytical continuation 
into a reliable tool for the study of steady-state properties of quantum impurity
systems using Han and Heary's formalism.

\section{Acknowledgments}
We acknowledge useful conversations with 
Jong Han, Sebastian Fuchs, Emanuel Gull, and Kurt Sch\"{o}nhammer.
A.D.\ further acknowledges the hospitality of 
the Center for Computation and Technology (CCT) at Louisiana State 
University and the financial
support by the German Academic Exchange Service (DAAD) through the PPP
exchange program. P.W. acknowledges support from SNF Grant PP002-118866.
M.J. acknowledges NSF grant DMR-0706379.

\appendix

\section{Derivation of the Kernel}
\label{sec:AppA}
Based on the argument given in section \ref{sec:shift_argument} we restrict ourselves to
the class of functions with positive imaginary part in the domain
$T^{C_\varepsilon}$, typically denoted as $H_+(T^{C_\varepsilon})$ in the
mathematical literature.
For a good overview of the concepts and terminologies used in the mathematical 
context see Ref.~\cite{bookVladimirov} and the first volume of Ref.~\cite{encyclMathSciences}.
 Vladimirov found the following generalization of
Herglotz-Nevanlinna representations to several complex variables \cite{Nevanlinna, Vladimirov1978}. 
It is essentially \cite{encyclMathSciences} the 
\par
\textbf{Theorem.} (Vladimirov, 1978/79) The following conditions for a
function $f\in H_+(T^{C})$ are equivalent for a cone $C \subset \mathbb{R}^m$
and $\mu(\underline x) := \Im f(\underline x)$:
\begin{enumerate}
\item The Poisson integral $P_C[\Dfrtl \mu]$ is pluriharmonic in $T^C$;
\item the function $\Im f(\underline z)$, $\underline z=\underline x + \imag \underline y\in T^C$, 
is represented by the Poisson formula
\begin{equation}
\Im f(\underline z) = P_C[\Dfrtl \mu](\underline z) + (\underline a,
\underline y),
\end{equation}
for some $a \in C^*$, where $C^*$ is the dual cone of $C$;
\item for all $\underline z^0\in T^C$, under the assumption that $C$ is
regular, the Schwarz representation
\begin{equation}
\begin{split}
f(\underline z) = &\imag \int_{\mathbb{R}^m} \mathcal{S}_C (\underline z -
\underline t, \underline z^0 - \underline t) \Dfrtl \mu (t) \\ & 
+ (\underline a, \underline z) + \underline b
\end{split}
\end{equation}
holds, with $b = b(\underline z^0) = \Re f(\underline z^0) - (\underline a,
\underline x^0)$.\hfill$\Box$
\end{enumerate}
Let us introduce the relevant mathematical terminology. 
A cone $C\subset
\mathbb{R}^m$ with vertex at zero is defined \cite{bookVladimirov} by the property that 
$y\in C \Rightarrow \forall \lambda > 0 : \lambda y \in C$.
Its dual cone $C^*:=\{\underline \xi \in \mathbb{R}^m\,|\, \forall \underline
x \in C : (\underline \xi, \underline x) \geq 0\}$. Here,
$P_C[\Dfrtl \mu](\underline z) = \int_{\mathbb{R}^m}\mathrm{d}^mx\, \mu(x)
\mathcal{P}_C(\underline z - \underline x)$ with the Poisson kernel
\begin{equation}
\mathcal{P}_C (\underline z) = 
\frac{|\mathcal{K}_C(\underline z)|^2}
{(2\pi)^m \mathcal{K}_C (2 \imag \underline y)},
\quad \underline z = \underline x + \imag \underline y
\label{eq:gendefpoisson}
\end{equation}
and the Cauchy kernel
\begin{equation}
\mathcal{K}_C(\underline z) = 
\int_{C^*} \mathrm{d}^m \xi\, \euler{\imag (\underline z,\underline \xi)}, \quad \underline z \in T^{C}.
\label{eq:gendefcauchy}
\end{equation}
We will not explicitly use the Schwarz kernel $\mathcal{S}$, the reader may
find it in Ref.~\cite{encyclMathSciences}.
A holomorphic mapping is said to be
biholomorphic iff it is one-to-one. Two domains $G,\tilde G$ are biholomorphically
equivalent iff a biholomorphic mapping $G\to \tilde G$ exists.
For the concept of
pluriharmonicity see introductory volumes of Ref.~\cite{encyclMathSciences}.
\par
In the case of $T^{C_\epsilon}$ we rewrite Eq.~\eqref{eq:defepsdomain} as
\begin{equation}
C_\epsilon = \bigcup_{\lambda\in(-\epsilon,\epsilon)}
\{(x_1,x_2)\in \mathbb{R}^2 | x_2 > 0 \wedge x_1 = \lambda x_2 \}.
\end{equation}
Hence, the dual cone 
\begin{equation*}
\begin{split}
C_\epsilon^* \,&=\, \bigcap_{\lambda \in (-\epsilon,\epsilon)}
\{
(\xi_1,\xi_2)\in\mathbb{R}^2 | \forall x_2>0 : \xi_1 \lambda x_2 +\xi_2 x_2 \geq 0
\}\\
\,&=\,
\{
(\xi_1,\xi_2)\in\mathbb{R}^2 | \xi_2 \geq 0 \wedge
\xi_1 \in [-\xi_2/\epsilon, \xi_2/\epsilon]
\}.
\end{split}
\end{equation*}
Evaluating the integrals $\int_{C_\epsilon^*}\mathrm{d}^m \xi =
\int_0^\infty \Dfrtl\xi_2 \int_{-\xi_2/\epsilon}^{\xi_2/\epsilon}
\Dfrtl{\xi_1}$
in \eqref{eq:gendefcauchy} yields
\begin{equation}
\mathcal{K}_{C_\epsilon}(\underline z) = 
-\frac{2}{\epsilon} \prod_{\mu=\pm 1} \frac{1}{z_2 - \mu z_1 / \epsilon}.
\end{equation}
Eq.~\eqref{eq:poissonkernelepsdomain} follows immediately from the 
definition \eqref{eq:gendefpoisson}.

In order to prove the validity of the representation 
\eqref{eq:kernel_Tceps} based on Vladimirov's theorem, we first determine $\underline
a = 0$ due to the boundedness of the Green's function.
Now we need to show that
the Poisson integral $P_{C_\epsilon}[\Dfrtl \mu]$ with respect to the 
measure $\mu(\underline x) = \Im f(\underline x)$ is pluriharmonic for 
all functions $f \in H_+(T^{C_\epsilon})$. 
Note that for the $m$-dimensional octant 
\begin{equation}
C_+^{(m)} := \mathbb{R}_+^m = \{(x_1,\dots, x_m) \in \mathbb{R}^m
| x_i > 0\}
\end{equation}
it was proven \cite{Vladimirov69,encyclMathSciences} that the Poisson kernel $\mathcal{P}_{C_+^{(m)}}$
is pluriharmonic for all functions $f\in H_+(T^{C_+^{(m)}})$.
Fortunately, as we restrict ourselves to $m=2$ in our application, all tubular cone
domains are known to be biholomorphically equivalent -- they are simply
connected through linear transformations.\par
To see the advantage more explicitly, we introduce the biholomorphism 
$M:T^{C_+^{(2)}}\to T^{C_\epsilon}$ given by the linear operation
\begin{equation}
M(\underline {\tilde z}) := M\cdot\underline {\tilde z} = 
\frac{1}{\sqrt{1+\epsilon^2}} 
\begin{pmatrix}
\epsilon  &  (1+\epsilon^2) / 2 \\
-1        &  (\epsilon + \epsilon^{-1})/2
\end{pmatrix}
\cdot \underline{\tilde z}.
\end{equation}
Obviously, 
\begin{equation}
M^{-1} = 
\frac{1}{\sqrt{1+\epsilon^2}} 
\begin{pmatrix}
 (\epsilon + \epsilon^{-1})/2  &  -(1+\epsilon^2) / 2 \\
1        &  \epsilon
\end{pmatrix}.
\end{equation}
We explicitly show that the kernel representation \eqref{eq:kernel_Tceps} for 
a function $f(\underline z)\in H_+(T^{C_\epsilon})$ may also be derived by
applying the corresponding Poisson kernel $\mathcal{P}_{C_+^{(2)}}$ for the tubular octant to the
corresponding function $\tilde f (\underline{\tilde z}) :=
f(M\underline{\tilde z}) \in H_+(T^{C_+^{(2)}})$ and transforming back to
$T^{C_+^{(2)}}$. Since the representation for $\tilde f$ is valid, we will
have shown explicitly that \eqref{eq:kernel_Tceps} is valid for all
$f\in H_+(T^{C_\epsilon})$.
\par
For this purpose it suffices to show that 
\begin{equation}
\mathcal {K}_{C_\epsilon}
(\underline z) = \mathcal {K}_{C_+^{(2)}} (M^{-1}\underline z),
\label{eq:toshow}
\end{equation} 
because then
$\mathcal {P}_{C_\epsilon}
(\underline z - \underline x) = \mathcal {P}_{C_+^{(2)}} (M^{-1}\underline z
- M^{-1}\underline x)$ and therefore 
$\mathcal {P}_{C_\epsilon}
(\underline z - M\underline{\tilde x}) = \mathcal {P}_{C_+^{(2)}} (M^{-1}\underline z
- \underline {\tilde x})$. We introduced the integration variables $\underline x$
and $\underline {\tilde x}$ of the Poisson integrals $P_{C_\epsilon}[\Dfrtl
\mu]$, $\mu(x) = \Im f(\underline x)$
and $P_{C_+^{(2)}}[\Dfrtl {\tilde \mu}]$, $\tilde \mu(\tilde x) = \tilde f(\tilde
x)$, respectively. Since $\det M = 1$, transforming $\tilde x\to x$ 
in $P_{C_+^{(2)}}$ then yields \eqref{eq:kernel_Tceps}.
\par
With a similar procedure as for $\mathcal{K}_{C_\epsilon}$ it is straightforward to show that 
\begin{equation}
\mathcal{K}_{C_+^{(2)}}(\underline {\tilde z}) = \frac{1}{\tilde z_1 \tilde
z_2}, \quad \underline {\tilde z} \in T^{C_+^{(2)}}.
\end{equation}
To finish the argument we verify that Eq.~\eqref{eq:toshow} holds by inserting
\begin{equation*}
\begin{split}
\mathcal{K}_{C_+^{(2)}} (M^{-1}\underline z)\,=\,&
(1+\epsilon^2)\cdot \left(\frac{\epsilon+\epsilon^{-1}}{2}z_1 - \frac{1+\epsilon^2}{2}
z_2\right)^{-1}\cdot \\
&\,\cdot \left(z_1 + \epsilon z_2\right)^{-1}.
\end{split}
\end{equation*}
Representations for any tubular cone domains in $\mathbb{C}^2$ are similarly
related due to the biholomorphic equivalence.
In particular, valid representations for $T^{C^s_\nu}$ are obtained easily.
For example, the Poisson kernel with respect to $T^{C^+_\nu}$ reads 
\begin{equation}
\mathcal{P} (\underline z) = 
\frac{1}{\pi^2} \prod_{\mu=\pm 1} 
\frac{y_2 - (\nu+\mu) y_1 /2}
{
(x_2 - \frac{\nu+\mu}{2} x_1)^2 + (y_2 - \frac{\nu+\mu}{2} y_1)^2
}
\label{eq:poissonkernelCplusNudomain}
\end{equation}
and could in principle be used for an enhanced continuation procedure
invoking data from all sectors of the complex space.

\end{document}